\documentstyle[12pt]{article}
\newcommand{\tresj}[6]{ \left( \begin{array}{ccc}
                               #1 & #2 & #3 \\
                               #4 & #5 & #6 
                             \end{array}
                        \right) } 
\newcommand{\seisj}[6]{ \left\{ \begin{array}{ccc}
                                #1 & #2 & #3 \\
                                #4 & #5 & #6 
                               \end{array}
                        \right\} } 
\newcommand{\nuevej}[9]{ \left\{ \begin{array}{ccc}
                                 #1 & #2 & #3 \\
                                 #4 & #5 & #6 \\
                                 #7 & #8 & #9 
                                \end{array}
                         \right\} } 
\newcounter{pepe}
{\end{eqnarray}%
\setcounter{equation}{\arabic{pepe}}%
}
\newcommand{\Jb}{{\cal J}}
\newcommand{\Mb}{{\cal M}}
\newcommand{\Wb}{{W}}
\newcommand{\np}{{\bf p}}
\newcommand{\nq}{{\bf q}}
\newcommand{\nr}{{\bf r}}

\newcommand{\nb}{{\bf b}}
\newcommand{\nw}{{\bf w}}
\newcommand{\nA}{{\bf A}}
\newcommand{\nB}{{\bf B}}
\newcommand{\nM}{{\bf M}}
\newcommand{\nX}{{\bf X}}
\newcommand{\nY}{{\bf Y}}
\newcommand{\nne}{{\bf e}}
\newcommand{\nsigma}{\mbox{\boldmath$\sigma$}}

\begin{document}
\begin{titlepage}
\mbox{}
\vspace*{2.5\fill}

\noindent
{\large 
{\bf Inclusive quasielastic scattering of polarized electrons \hfill\\
     from polarized nuclei$^*$}
}

\vspace{1\fill}

\noindent
J.E. Amaro$^{1,2}$, J.A. Caballero$^{3,4}$, T.W. Donnelly$^1$ and 
 E. Moya de Guerra$^3$ \hfill\\[3mm]
$^1$ Center for Theoretical Physics, Laboratory for Nuclear
     Science and Dept. of Physics, Massachusetts Institute of 
     Technology, Cambridge, MA 02139, U.S.A. \\[2mm]
$^2$ Departamento de F\'{\i}sica Moderna, Universidad de Granada,\hfill\\
      Granada 18071, Spain \hfill\\[2mm]
$^3$ Instituto de Estructura de la Materia, Consejo superior 
     de Investigaciones Cient\'{\i}ficas, Serrano 123, Madrid 28006, Spain 
     \hfill\\[2mm]
$^4$ Departamento de F\'{\i}sica At\'omica, Molecular y Nuclear, 
     Universidad de Sevilla, Aptdo 1065, Sevilla 41080, Spain.

\vspace{2\fill}

\begin{quote}
{\small
{\bf Abstract}
\vspace{0.3cm}\par\penalty10000

The inclusive quasielastic response functions that appear in the scattering 
of polarized electrons from polarized nuclei are computed and analyzed
for several closed-shell-minus-one nuclei with special attention paid to 
$^{39}$K. Results are presented using two models for the ejected nucleon 
--- when described by a distorted wave in the continuum shell model or by 
a plane wave in PWIA with on- and off-shell nucleons. Relativistic effects 
in kinematics and in the electromagnetic current have been incorporated
throughout. Specifically, the recently obtained expansion of the 
electromagnetic current in powers only of the struck nucleon's momentum is 
employed for the on-shell current and the effects of the first-order terms 
(spin-orbit and convection) are compared with the zeroth-order (charge 
and magnetization) contributions. The use of polarized inclusive quasielastic 
electron scattering as a tool for determining near-valence nucleon 
momentum distributions is discussed.
}
\end{quote}

\vfill
$^*$ This work is supported in part by funds provided by the U.S. 
Department of Energy (D.O.E.) under cooperative agreement 
\#DE-FC01-94ER40818 (T.W.D. and J.E.A.), in part by DGICYT (Spain) under 
Contract 
Nos. PB92/0021-C02-01 and PB92-0927 and the Junta de Andaluc\'{\i}a (Spain)
and in part by NATO Collaborative Research Grant \#940183.

\noindent MIT/CTP\#2549 \hfill July 1996

\end{titlepage}

\newpage
\section{Introduction}

Inclusive quasielastic electron scattering is well-known as a way to probe
the charge and current densities of individual nucleons within the nucleus.
In this regime, at high momentum transfers $q$ and for energy transfers 
$\omega\cong\sqrt{q^2+M^2}-M$, where $M=$ nucleon mass, single-nucleon 
knockout is expected to be dominant. The inclusive unpolarized reaction
$A(e,e')$ can be analyzed relatively directly in terms of the hadronic 
responses labeled $L$ and $T$ for longitudinal (charge) and transverse 
(current) contributions, respectively. 
To date the main attention has been on the completely unpolarized 
reaction, but now the study of  electron scattering of polarized
electrons from polarized nuclei is becoming feasible,
due to recent technological advances in the development
of polarized targets and high-duty-factor electron beams. 
The amount of nuclear structure information which is obtainable
in such instances can be considerably richer than in the
unpolarized case in that more responses of wider diversity can be studied 
\cite{Don86}. While in other investigations devoted to polarization in
quasielastic electron scattering the emphasis has been placed on the exclusive
coincidence observables (see, for example, refs.~\cite{Ras89}, 
\cite{Bof93}, \cite{Bof88} \cite{Cab93} 
and \cite{Gar95}), in this work our focus is on inclusive scattering with 
polarization degrees of freedom. Most such studies have been restricted to 
$^2{\vec {\rm H}}$ and $^3{\vec {\rm He}}$, and include recent experimental 
asymmetry measurements \cite{Gao94,Han95}; however, in this paper we restrict 
our attention entirely to polarization observables in (inclusive) quasielastic 
electron scattering from medium and heavy nuclei. 

The organization of the paper is the following: in sect.~\ref{Formalism} we
present the formalism, centering the discussion around several
issues. (1) We briefly review the essential expressions for the general 
nuclear polarized electromagnetic response functions
based on the material presented in ref.~\cite{Don86} and establish
the notation that follows. (2) We apply the above formalism to the particular 
case of one-hole nuclei in the continuum shell model (CSM) (see 
ref.~\cite{Ama94}), with some details concerning the analytical sum over the
angular momenta of the final hadronic state discussed in appendix A and 
concerning the Coulomb multipoles for the spin-orbit charge density placed 
in appendix B. While for the sake of brevity our focus is placed on 
one-proton-hole cases, the approach is straightforwardly applied to 
one-neutron-hole nuclei as well as to nuclei having a single nucleon above a 
closed shell. In the present work when using the CSM we take the 
mean field in initial and final states to be the same, thereby 
maintaining the orthogonality of the single-particle wave functions. 
Clearly at high outgoing nucleon energies this exaggerates the final-state 
interactions (FSI) and thus we contrast this approach with the 
Plane Wave Impulse Approximation (PWIA). Accordingly, 
(3) we summarize the essence of the PWIA for both on- and off-shell 
single-nucleon currents. Indeed, 
one of our motivations in the present work is to explore the evolution 
from the low-energy regime where final-state interactions (FSI) are 
important to higher energies where they are not and where the PWIA is 
expected to be valid. Intercomparisons of the CSM and PWIA results will also 
serve in explaining many of the features of the spin observables. In 
discussing the PWIA we obtain the polarized responses for the case of a 
one-hole nucleus (again, with some details of the calculation presented in 
appendix C) and introduce a version of the model where the (usually 
off-shell) relativistic single-nucleon current is replaced by the on-shell 
form used in recent work with promising results \cite{Ama96}. 
This current involves an expansion only in powers of $\eta\equiv p/M$,
with $p$ the bound nucleon momentum, and importantly not in $\kappa\equiv 
q/2M$ or $\lambda\equiv \omega/2M$.

In sect.~\ref{Results} we present the results for the response functions of 
a few selected one-proton-hole nuclei. 
The discussions in this section start with a re-examination of the elastic 
scattering of polarized electrons from a polarized proton at rest, where we 
know the exact answer, in order to provide some insight into the cases of 
polarized protons in $1s_{1/2}$, $1p_{1/2}$ and $2s_{1/2}$ shells. The 
results in these cases are contrasted with those for $1p_{3/2}$ and $1d_{3/2}$ 
polarized protons, the last being the case that is then studied in more 
detail, namely, the nucleus $^{39}$K. In the course of presenting our results 
we study the effects due to distortion in the ejected nucleon wave function 
caused by the real nuclear mean field, as well as the effects of the 
first-order ($\eta$-dependent) terms in the nuclear current; second-order 
terms are also briefly discussed here. On- and off-shell versions of the 
PWIA are compared with each other and with the $\eta$-dependent effects. 
We discuss the sensitivity of the polarized responses to the actual orbit 
containing the polarized nucleon and how this has the potential to be a 
powerful new tool for determining the near-valence-shell nucleon momentum 
distribution. Finally, in sect.~\ref{Concl}, we draw our conclusions.

\section{Formalism}\label{Formalism}

\subsection{Polarized cross section}\label{Polxsec}

We consider a nucleus $A$  in an initial state
polarized in direction $\Omega^*=(\theta^*,\phi^*)$,
denoted $|A\rangle=|i(\Omega^*)\rangle$, from which a polarized
electron with helicity $h$ is scattered, leaving the nucleus 
in a final state $|f\rangle$ lying in the continuum.  In the
approximation that the electron is described by plane waves, the
inclusive cross section in the laboratory system may be written in
terms of six nuclear response functions \cite{Don86}:
\begin{equation} 
\frac{{\rm d}^2\sigma}{{\rm d}\epsilon'_e \,{\rm d}\Omega'_e} 
= \sigma_{\rm M} 
\left[\sum_K v_K R^K + h \sum_{K'}v_{K'}R^{K'}\right], \label{eq1}
\end{equation}
where $Q^\mu=(\omega,{\bf q})$ is the four-momentum transfer, 
$\epsilon'_e$ and $\Omega'_e$ are the  energy and angles of the
final electron,
$\sigma_{\rm M}$ is  the Mott cross section and
the summation indices are $K=L,T,TL,TT$ and $K'=T',TL'$. 
If the extreme relativistic limit is assumed for the electron and
we refer to a coordinate system with the $z$-axis along ${\bf q}$
and the $x$-axis in the scattering plane, the kinematic factors 
$v_K$ and $v_{K'}$ can be written as
\begin{eqnarray}
v_L=\left({\frac{Q^{2}}{q^{2}}}\right)^2, &&
v_T=\tan^{2}\frac{\theta_e}{2}-\frac{Q^{2}}{2q^{2}}, \\
v_{TL}=\frac1{\sqrt{2}}\frac{Q^2}{q^2}
       \sqrt{\tan^2\frac{\theta_e}{2}-\frac{Q^2}{q^2}}, &&
v_{TT}=\frac{Q^2}{2q^2}, \\
v_{T'}=\tan\frac{\theta_e}{2}
        \sqrt{\tan^2\frac{\theta_e}{2}-\frac{Q^2}{q^2}}, &&
v_{TL'}=\frac1{\sqrt{2}}\frac{Q^2}{q^2}\tan\frac{\theta_e}{2}.
\end{eqnarray}
Here we use $Q_{\mu}Q^{\mu}\equiv Q^2= \omega^2-q^2<0$ 
and  $\theta_e$ is the electron scattering angle. 
The inclusive nuclear responses $R^{K},R^{K'}$ 
are the components 
\begin{eqnarray}
R^L=W^{00},&& 
R^T=W^{xx}+W^{yy} \label{componentes1}\\
R^{TL}=\sqrt{2}(W^{0x}+W^{x0}), && 
R^{TT}=W^{yy}-W^{xx}\label{componentes2}\\
R^{T'}=i(W^{xy}-W^{yx}), &&
R^{TL'}=i\sqrt{2}(W^{0y}-W^{y0}) \label{componentes3}
\end{eqnarray}
of the nuclear electromagnetic tensor defined by
\begin{equation}
W^{\mu\nu}=\sum_f\delta(E_f-E_i-\omega)
           \langle f|\hat{J}^{\mu}({\bf q})|i\rangle^*
           \langle f|\hat{J}^{\nu}({\bf q})|i\rangle.
\end{equation}
Here the sum runs over all the final nuclear states $|f\rangle$
with energy $E_f=E_i+\omega$ and
\begin{equation}
\hat{J}^{\mu}({\bf q})=\int\,d^3r\,\,{\rm e}^{i\bf q\cdot r}
                       \hat{J}^{\mu}({\bf r})
\end{equation}
is the Fourier transform of the nuclear electromagnetic operator.

\subsection{Multipole analysis of the polarized responses}

We now consider the case when the initial nuclear state of spin $J_i$ 
is fully polarized in the direction $\Omega^*=(\theta^*,\phi^*)$,
where the spherical coordinates for the polarization axis are taken with
respect the coordinate system described above (see the developments in 
refs.~\cite{Don86} and \cite{Ras89} for details). Using rotation matrices 
the initial nuclear wave function can be expressed in terms of states 
$|J_iM_i\rangle$ for which the axis of quantization is {\bf q}:
\begin{equation}
|i\rangle = |J_iJ_i(\Omega^*)\rangle=
          \sum_{M_i}{\cal D}^{(J_i)}_{M_iJ_i}(\Omega^*)|J_iM_i\rangle.
\end{equation}
We must sum over all the final nuclear states and thus we may also use a basis
of final states with good angular momentum having {\bf q} as the axis of 
quantization, $|f\rangle=|J_fM_f\rangle$. The charge and transverse current 
operators may be developed as sums of multipoles in the usual way:
\begin{eqnarray}
\hat{\rho}({\bf q}) &=& \sqrt{4\pi}\sum_J i^J[J] 
                        \hat{M}_{J0}(q)\\
\hat{J}_{M}({\bf q})&=& -\sqrt{2\pi}\sum_J i^J[J]
                \left[ \hat{T}^{el}_{JM}(q)+M\hat{T}^{mag}_{JM}(q)\right],
\end{eqnarray}
where $M$ is the index of the spherical components and $\hat{M}_{JM}$,
$\hat{T}^{el}_{JM}$ and $\hat{T}^{mag}_{JM}$ are the usual Coulomb (CJ), 
electric (EJ) and magnetic (MJ) multipoles. Throughout we use the notation 
$[J]\equiv\sqrt{2J+1}$ for any angular momentum variable $J$.

Using the above expansion, the nuclear tensor can be written as a sum
of factors of the kind
\begin{equation}
B^{M'M}_{J'J} =  \sum_{M_fM_iM'_i}
                 {\cal D}^{(J_i)*}_{M_iJ_i}
                 {\cal D}^{(J_i)}_{M'_iJ_i}
                 \langle J_fM_f|\hat{T}'_{J'M'}|J_iM_i\rangle^*
                 \langle J_fM_f|\hat{T}_{JM}|J_iM'_i\rangle,
\end{equation}
where $T$ and $T'$ are any of the $C$, $E$ or $M$ operators.
The dependence on the polarization axis in
the product of two rotation matrices can be developed in spherical harmonics
\begin{equation}
{\cal D}^{(J_i)*}_{M_iJ_i}{\cal D}^{(J_i)}_{M'_iJ_i} =
\sqrt{4\pi}\sum_{\cal JM}(-1)^{J_i+M_i+{\cal J}}
\tresj{J_i}{J_i}{\cal J}{-M_i}{M'_i}{\cal M}
f^{i}_{\cal J} Y_{\cal JM}(\Omega^*),
\end{equation}
where $f^i_{\cal J}$ is the Fano tensor for 100\% polarization, 
$f^i_{\cal J}=\langle J_iJ_iJ_i-J_i|{\cal J}0\rangle$.
Then using the Wigner-Eckart theorem and suming over third
components, one obtains the following dependence involving
the polarization angles and the reduced matrix elements of
the diferent multipole operators (see ref.~\cite{Don86} for details):
\begin{eqnarray}
B^{M'M}_{J'J} &=& \sum_{\cal JM}(-1)^{J_i+J_f+M+\cal M}
                 [{\cal J}]\left[\frac{\cal(J-M)!}{\cal(J+M)!}\right]^{1/2}
                 \tresj{J'}{J}{\cal J}{M'}{-M}{\cal M}\nonumber\\
              && \times\seisj{J'}{J}{\cal J}{J_i}{J_i}{J_f} 
                 f^i_{\cal J}P^{\cal M}_{\cal J}(\cos\theta^*)
                 {\rm e}^{i{\cal M}\phi^*}                 
                 \langle J_f\|\hat{T}'_{J'}\|J_i\rangle^*
                 \langle J_f\|\hat{T}_{J}\|J_i\rangle,
\end{eqnarray}
where $P^{\cal M}_{\cal J}(\cos\theta^*)$ is an associated
Legendre function (we use the conventions of ref.~\cite{Edmonds}).

Now taking the appropiate components of the nuclear tensor  and defining
the (real) Coulomb, electric and magnetic multipoles
\begin{eqnarray}
T_{CJ} &\equiv& \langle J_f\|\hat{M}_{J}(q)\|J_i\rangle \label{mb1}\\
T_{EJ} &\equiv& \langle J_f\|\hat{T}^{el}_{J}(q)\|J_i\rangle\\
T_{MJ} &\equiv& \langle J_f\|i\hat{T}^{mag}_{J}(q)\|J_i\rangle, \label{mb2}
\end{eqnarray}
we arrive at the following equations for the polarized quasielastic
responses \cite{Don86}:
\begin{eqnarray}
R^L    &=& 4\pi\sum_{\cal J}P^+_{\cal J}P_{\cal J}(\cos\theta^*)
           f^i_{\cal J}{W}^L_{\cal J}(q,\omega) \label{ea5} \\
R^T    &=& 4\pi\sum_{\cal J}P^+_{\cal J}P_{\cal J}(\cos\theta^*)
           f^i_{\cal J}{W}^T_{\cal J}(q,\omega) \\
R^{TL} &=& 4\pi\sum_{\Jb\ge2}P^+_{\cal J}P^1_{\cal J}(\cos\theta^*)
           \cos\phi^*f^i_{\cal J}{W}^{TL}_{\cal J}(q,\omega) \\
R^{TT} &=& 4\pi\sum_{\Jb\ge 2}P^+_{\cal J}P^2_{\cal J}(\cos\theta^*)
           \cos2\phi^*f^i_{\cal J}{W}^{TT}_{\cal J}(q,\omega) \\
R^{T'} &=& 4\pi\sum_{\cal J}P^-_{\cal J}P_{\cal J}(\cos\theta^*)
           f^i_{\cal J}{W}^{T'}_{\cal J}(q,\omega) \\
R^{TL'}&=& 4\pi\sum_{\cal J}P^-_{\cal J}P^1_{\cal J}(\cos\theta^*)
           \cos\phi^*f^i_{\cal J}{W}^{TL'}_{\cal J}(q,\omega), \label{ea6}
\end{eqnarray}
where the parity projectors $P_{\cal J}^{\pm}\equiv[1+(-1)^{\cal J}]/2$
have been used. The above expressions show that several
sets of measurements with different polarization angles $\Omega^*$
in principle lead to the separation of the reduced nuclear responses
${W}_{\cal J}^K(q,\omega)$; these are functions only of
the energy and momentum transfer and are given by the following: for 
${\cal J}=$ even
\begin{eqnarray}
{W}^L_{\cal J} &=& \sum_f\delta(E_f-E_i-\omega)
                        \sum_{J'J}
                        \Lambda_{fi}(J',J,\Jb,0,0,0)
                        \xi^+_{J'J}T_{CJ'}T_{CJ} \label{ea1} \\
{W}^T_{\cal J} &=& -\sum_f\delta(E_f-E_i-\omega)
                        \sum_{J'J}
                        \Lambda_{fi}(J',J,{\cal J},1,-1,0)\nonumber\\
                    & & \times [\xi^+_{J'J}(T_{EJ'}T_{EJ}+T_{MJ'}T_{MJ})+
                         \xi^-_{J'J}(T_{EJ'}T_{MJ}-T_{MJ'}T_{EJ})] \\
{W}^{TL}_{\cal J} &=& 2\sqrt{2}\sum_f\delta(E_f-E_i-\omega)
                        \sum_{J'J}
                        \Lambda_{fi}(J',J,{\cal J},0,1,-1)\nonumber\\
                    & & \times T_{CJ'}(\xi^+_{J'J}T_{EJ}-\xi^-_{J'J}T_{MJ})\\
{W}^{TT}_{\cal J} &=& -\sum_f\delta(E_f-E_i-\omega)
                        \sum_{J'J}
                        \Lambda_{fi}(J',J,{\cal J},1,1,-2)\nonumber\\
                    & & \times [\xi^+_{J'J}(T_{EJ'}T_{EJ}-T_{MJ'}T_{MJ})-
                         \xi^-_{J'J}(T_{EJ'}T_{MJ}+T_{MJ'}T_{EJ})], 
\end{eqnarray}
whereas for ${\cal J}=$ odd, ${W}^T_{\cal J} \rightarrow 
{W}^{T'}_{\cal J}$ and ${W}^{TL}_{\cal J} \rightarrow 
-{W}^{TL'}_{\cal J}$ (see ref.~\cite{Don86}). We use the notation
\begin{eqnarray}
\Lambda_{fi}(J',J,{\cal J},M',M,{\cal M})
          & \equiv & [J][J'][\Jb](-1)^{J_i+J_f}
                \left[\frac{(\Jb-|\Mb|)!}{(\Jb+|\Mb|)!}\right]^{1/2}
                \nonumber\\
          &   & \times\tresj{J'}{J}{\cal J}{M'}{M}{\Mb}
                \seisj{J'}{J}{\cal J}{J_i}{J_i}{J_f} \label{ea2}
\end{eqnarray}
and
\begin{equation}
\xi_{J'J}^+\equiv (-1)^{(J'-J)/2}P^+_{J'+J}, \kern 1cm
\xi_{J'J}^-\equiv (-1)^{(J'-J+1)/2}P^-_{J'+J}.
\end{equation}
These equations are general for any nucleus. 
In the next section we apply them to the particular 
case of a one-hole nucleus.

\subsection{Polarized responses for a one-hole nucleus}

In this work we consider closed-shell-minus-one nuclei where the
initial nuclear state is obtained as a hole in a closed-shell core $|C\rangle$:
\begin{equation}
|A\rangle=|i^{-1}(\Omega^*)\rangle=
\sum_{m_i}{\cal D}^{(j_i)}_{m_ij_i}(\Omega^*)b_{i,m_i}^{\dagger}|C\rangle,
\end{equation}
where $|i\rangle=|n_i,l_i,j_i\rangle$ is a single-particle
state occupied in the core and $b_{i,m_i}^{\dagger}$ is the creation
operator for a hole, $b_{i,m_i}^{\dagger}=(-1)^{j_i+m_i}a_{i,-m_i}$. 
Here we follow the convention of using lower case letters $j_i$ for half-integer 
angular momenta. 
As in the present work we only consider the one-body piece 
of the electromagnetic nuclear current, the
interaction with the virtual photon gives rise only to particle-hole
(p-h) excitations. Thus the final nuclear states are described by
\begin{equation}
|f\rangle=|p,(h^{-1},i^{-1})J_B;j_f\rangle.
\end{equation}
 Here $|h\rangle=|n_h,l_h,j_h\rangle$ is another (bound) single-particle 
state in the core, while
$|p\rangle=|\epsilon_p,l_p,j_p\rangle$ is a particle in the continuum.
The residual nucleus is a two-hole nucleus
$|B\rangle=|(h^{-1},i^{-1})J_B\rangle$ with total angular
momentum $J_B$, and it is coupled with the outgoing particle 
$|p\rangle$ to a total angular momentum $j_f$. In the present work the 
wave functions of the single-particle states are obtained
using a mean-field potential of
Woods-Saxon type, for both negative (bound) and positive 
(continuum) energies.  More details on this aspect of the calculation,
including the values of the potential parameters,
can be found in refs.~\cite{Ama94} and \cite{Ama96}. 

The sum over final states to obtain the inclusive responses now
runs over all the holes $h$, particles $p$, and angular momenta $J_B$
and $j_f$. In appendix A we show how the sums over $J_B$
and $j_f$ can be performed using Racah algebra. The final
reduced responses can be written in terms of the single-particle 
reduced matrix elements taken between particle and hole states
\begin{eqnarray}
t_{CJ} &\equiv& \langle p\|M_{J}(q)\|h\rangle\\
t_{EJ} &\equiv& \langle p\|T^{el}_{J}(q)\|h\rangle\\
t_{MJ} &\equiv& \langle p\|iT^{mag}_{J}(q)\|h\rangle,
\end{eqnarray}
using lower-case letters to distinguish the particle-hole multipole 
matrix elements from their many-body counterparts in 
eqs.~(\ref{mb1}--\ref{mb2}), 
and yield the following expressions for the reduced responses:
\begin{eqnarray}
{W}^L_{\cal J} &=& \delta_{\Jb 0}[j_i]
                        \sum_{ph}\delta(\epsilon_p-\epsilon_h-\omega)
                        \sum_Jt_{CJ}^2\nonumber\\
                    & &\mbox{}-
             \sum_{ph}\delta_{hi}\delta(\epsilon_p-\epsilon_h-\omega)
                        \sum_{J'J}
                        \Lambda_{ph}(J',J,{\cal J},0,0,0)
                        \xi^+_{J'J}t_{CJ'}t_{CJ} \\
{W}^T_{\cal J} &=& \delta_{\Jb 0}[j_i]
                        \sum_{ph}\delta(\epsilon_p-\epsilon_h-\omega)
                        \sum_J(t_{EJ}^2+t_{MJ}^2)\nonumber\\
                    & &\mbox{}+(-1)^{\Jb}
             \sum_{ph}\delta_{pi}\delta(\epsilon_p-\epsilon_h-\omega)
                        \sum_{J'J}
                        \Lambda_{ph}(J',J,{\cal J},1,-1,0)\nonumber\\
                    & & \times [\xi^+_{J'J}(t_{EJ'}t_{EJ}+t_{MJ'}t_{MJ})+
                         \xi^-_{J'J}(t_{EJ'}t_{MJ}-t_{MJ'}t_{EJ})] \\
{W}^{TL}_{\cal J} 
                    &=& -2\sqrt{2}(-1)^{\Jb}
             \sum_{ph}\delta_{hi}\delta(\epsilon_p-\epsilon_h-\omega)
                        \sum_{J'J}
                        \Lambda_{ph}(J',J,{\cal J},0,1,-1) \nonumber\\
                    & & \times t_{CJ'}(\xi^+_{J'J}t_{EJ}-\xi^-_{J'J}t_{MJ})\\
{W}^{TT}_{\cal J} 
                    &=&(-1)^{\Jb}
             \sum_{ph}\delta_{pi}\delta(\epsilon_p-\epsilon_h-\omega)
                        \sum_{J'J}
                        \Lambda_{ph}(J',J,{\cal J},1,1,-2)\nonumber\\
                    & & \times [\xi^+_{J'J}(t_{EJ'}t_{EJ}-t_{MJ'}t_{MJ})-
                         \xi^-_{J'J}(t_{EJ'}t_{MJ}+t_{MJ'}t_{EJ})] \\
{W}^{T'}_{\cal J} &=& (-1)^{\Jb}
             \sum_{ph}\delta_{pi}\delta(\epsilon_p-\epsilon_h-\omega)
                        \sum_{J'J}
                        \Lambda_{ph}(J',J,{\cal J},1,-1,0)\nonumber\\
                    & & \times [\xi^+_{J'J}(t_{EJ'}t_{EJ}+t_{MJ'}t_{MJ})+
                         \xi^-_{J'J}(t_{EJ'}t_{MJ}-t_{MJ'}t_{EJ})] \\
{W}^{TL'}_{\Jb} &=& -\Wb^{TL}_{\Jb}.
\end{eqnarray}
Note that although formally 
${W}^{TL'}_{\Jb} = -\Wb^{TL}_{\Jb}$, in practice they are
quite different because of the different ${\cal J}$-values
(the same comment applies to ${W}^{T'}_{\cal J}$
versus ${W}^{T}_{\cal J}$). 
We shall see in the next section that an explicit calculation 
in PWIA shows that actually the primed and unprimed responses have quite 
different analytical forms.

Some comments about the above equations are now in order.
\begin{enumerate}
\item The polarized responses for the case of one particle above a closed 
shell are simply related to these one-hole results:
\begin{equation}
\left( {W}_{\cal J}^K \right)_{\rm particle}=-(-1)^{\cal J}
\left( {W}_{\cal J}^K \right)_{\rm hole}. \label{eqn42}
\end{equation}

\item The first terms in the $L$ and $T$ responses only
contribute for $\Jb=0$ and they give rise to
the unpolarized reponses of the closed-shell nucleus $|C\rangle$.
In fact, for $\Jb=0$ we have the Fano tensor $f_0^i = 1/[j_i]$ and
the Legendre function $P_0(\cos\theta^*)=1$, leading to
the following piece in the longitudinal response
\begin{eqnarray}
4\pi\sum_{\Jb}P_{\Jb}(\cos\theta^*)f^i_{\Jb}\delta_{\Jb 0}[j_i]
    \sum_{phJ}\delta(\epsilon_p-\epsilon_h-\omega)
    t_{CJ}^2\nonumber\\
=4\pi\sum_{phJ}\delta(\epsilon_p-\epsilon_h-\omega)t_{CJ}^2
\end{eqnarray}
and a similar term  for the transverse response.

\item The second term of $\Wb_{\Jb}^L$ or $\Wb_{\Jb}^T$
only contributes for $h=i$ and corresponds to
the reponse of the hole; it must be subtracted
from the total response of the core to obtain the total response
of the nucleus. For example, for $\Jb=0$ 
we have $\Lambda_{pi}(J,J,0,0,0,0)=1/[j_i]$ and thus obtain
for the $\Wb_0^L$-response
\begin{equation}
\Wb_0^L = \sum_{phJ}\delta(\epsilon_p-\epsilon_h-\omega)[j_i]
            \left(1-\frac{\delta_{hi}}{2j_i+1}\right)t_{CJ}^2,
\end{equation}
that is, the total unpolarized response of the core minus the response of
the hole.

\item For the polarized responses TL, TT, T' and TL' only the
shell containing the hole contributes. 
Note that when one has a single particle outside a closed-shell core then 
these contributions behave as shown in eq.~(\ref{eqn42}).

\item For the electromagnetic current operators we use the 
non-relativistic reduction whose expressions are given below
(eqs.~(\ref{current-spin-matrix}), 
(\ref{rho}-\ref{j2})). The time component is the sum of charge
plus spin-orbit operators, while the spatial component is 
the sum of  magnetization plus  convection operators.
The reduced matrix elements of the  (charge) Coulomb, and
(magnetization and convection) electric and magnetic operators  
are given in ref.~\cite{Ama96}. 
Note that, although we have maintained the traditional names for 
these terms, actually they differ slightly from the traditional 
charge, magnetization and
convection operators. In the present work we include the spin-orbit
operator which is usually not included in non-relativistic 
electron scattering calculations, since it is assumed to be
small. As we shall see, the spin-orbit contributions do turn out to be 
small for the unpolarized responses, but are
important for some of the polarized ones (especially the TL response) 
at the values of the momentum transfer used in this work.
The reduced matrix elements of the Coulomb operator for the
spin-orbit term are computed in appendix B. 

\item In all the calculations
that follow we employ relativistic kinematics. This means that 
from the (non-relativistic) energy of 
the emitted nucleon $\epsilon_p=\epsilon_h+\omega$
 we compute the momentum $\np'$
of the particle in PWIA from the relativistic energy-momentum relation
\begin{eqnarray}
p'{}^2 
& = &(M+\epsilon_p)^2-M^2 = \epsilon_p(\epsilon_p+2M)\nonumber\\
& = & 2M\left[\omega\left(1+\frac{\omega}{2M}\right)
               +\epsilon_h
               +\epsilon_h\frac{2\omega+\epsilon_h}{2M}
         \right]. \label{prel}
\end{eqnarray}
When the final-state interactions are present ({\it i.e.,} in the CSM)
we solve the equivalent Schr\"odinger equation 
with eigenvalue $\epsilon_p(1+\epsilon_p/2M)$. Note 
that this procedure is equivalent to
solving a Klein-Gordon equation 
$(-\nabla^2+M^2+2MV)\psi = E_p^2\psi $,
with $V$ the Woods-Saxon potential and $E_p=\epsilon_p+M$ the
relativistic energy of the particle. 
An alternative, convenient prescription to implement the above
relativistic kinematics was employed
in ref.~\cite{Ama96} (see also ref.~\cite{Alb90}), that of making the 
replacement $\omega\rightarrow \omega(1+\omega/2M)$ 
in the non-relativistic nuclear responses, but not in the 
single-nucleon form factors (where the asymptotic momentum is computed as
$p'{}^2=2M(\omega+\epsilon_h)$). Clearly this is equivalent to the above 
equation (\ref{prel}) provided that the last term 
$\epsilon_h(2\omega+\epsilon_h)/2M$ can be neglected, which is a good
approximation at the energies considered in this work. 

\item The sums over multipoles $J,J'$ are infinite and, in practice, we must
sum all multipoles until convergence is reached. The number of
multipoles needed grows with the mass number $A$ and 
with the momentum transfer, as detailed in table~2 of 
ref.~\cite{Ama96}. Indeed, in this work we were able to test the 
degree of convergence by examining the particular
case where $V=0$ for the final state (but not for
the initial) using two different approaches, with the multipole expansion 
and with the factorization procedure in PWIA --- the results obtained were 
indistinguishable.

\end{enumerate}

\subsection{Polarized responses in PWIA}\label{PolPWIA}

The multipole expansions introduced in the last section may therefore
also be applied when we make the approximation in which the
particle $p$ is a free (non-interacting) wave function and so described as a
plane wave function or, in a multipole analysis, a spherical Bessel 
function. Additional reasons for considering the plane-wave outgoing nucleon 
approximation, beyond being a test of convergence, are the following.
First, we want to analyze the effects arising from the distortion of the ejected
particle due to the mean field in the final state and compare these 
with the results obtained with a pure plane
wave. Note that in our treatment the distortion is caused by a
{\em real} mean-field potential. Therefore there is no absorption due
to inelastic processes in the exit channel. The reduction of flux
can be treated at least approximately {\em a posteriori} by using a 
phenomenological model of final-state interaction that takes care   
of those effects through the introduction of a self-energy for the particle-hole
propagator.  Alternatively, these effects could also be
simulated by a phenomenological
 optical potential in the exit channel, although
this raises the issue of lack of orthogonality in the wave functions.
The study of such additional FSI effects in the polarized
responses is left for a forthcoming paper and here we only deal
with  pure distortion effects
caused by the nuclear mean field.

Second, for the values of the momentum transfer
($q \sim 500$ MeV/c) that we are considering here 
the energy of the outgoing nucleon  in the quasielastic region
is very high in comparison with the energy of the bound nucleons.
This is especially true in the case of the polarization-dependent
responses, where only the outer  shell contributes.
Thus the approximation of plane waves for the ejected nucleon
should give reasonable results. The advantage of the PWIA approach
is that we can perform part of the calculation analytically 
and can factorize the responses into a product of two
terms, the first containing the single-nucleon responses and
the second the spectral function  of the polarized nucleus. 
This later approximation is very useful when attempting to disentagle
contributions from different terms in the nuclear current
operator (spin-orbit, convection, higher-order momentum-dependent 
terms,\ldots).

\subsubsection{Factorization of the responses}

We begin with the exclusive process $\vec{A}(\vec{e},e'N)B$
in which the ejected nucleon has momentum $\np'$ and the residual
nucleus is left in the state $|B\rangle$. The exclusive cross section
for such a process is given by ({\it cf.} eq.~(\ref{eq1}))
\begin{equation} 
\frac{{\rm d}^3\sigma}{{\rm d}\epsilon'_e \,{\rm d}\Omega'_e\,{\rm d}\Omega'} 
= \sigma_{\rm M} 
\left[\sum_K v_K {\cal R}_B^K + h \sum_{K'}v_{K'}{\cal R}_B^{K'}\right],
\end{equation}
where ${\cal R}_B^{K,K'}$ are the exclusive responses. 
The inclusive responses $R^K$ are obtained by integration over
the nucleon angles and summing over the states $|B\rangle$ 
of the residual nuclear system
\begin{equation}
R^{K,K'}=\sum_B\int {\rm d}\Omega'\,{\cal R}_B^{K,K'}(\Omega)
        =\sum_B R_B^{K,K'}.
\end{equation}
The inclusive cross section is given again by eq.~(\ref{eq1}). 
The exclusive responses are  combinations of the exclusive
hadronic tensor in PWIA
\begin{equation}
W_B^{\mu\nu}=Mp'\sum_{s'M_B}
\langle \np's'B|J^{\mu}(\nq)|A\rangle^*
\langle \np's'B|J^{\nu}(\nq)|A\rangle, \label{tenns}
\end{equation}
where $\langle \np's'B\rangle$ corresponds to a final state having an 
on-shell plane-wave nucleon with momentum $\np'$ and spin projection $s'$, 
together with the daughter nucleus in some state labeled $B$.

The details of proceeding in the usual approach to spin-dependent PWIA 
studies with off-shell single-nucleon currents $CC1$, $CC2$, {\it etc.,} 
are given in refs.~\cite{Cab93,Cab94, Deforest} and will not be repeated 
here. Instead, 
here we specialize the formalism to the situation where we take the PWIA with 
an on-shell current, specifically, the one developed in ref.~\cite{Ama96}. 
In the next section results are presented for both on- and off-shell cases. 
Proceeding with the former, the coordinate-space plane-wave nucleon's wave 
function is given by
\begin{equation} \label{onda-plana}
\langle \nr|\np's'\rangle=(2\pi)^{-3/2}{\rm e}^{i\np'\cdot\nr}\chi_{s'}
\end{equation}
with energy $E_{p'}=\omega+M_A-E_B=\sqrt{p'{}^2+M^2}$. Here only 
two-component spin spinors occur, since the lower-component information 
that is usually explicit in the PWIA has been incorporated in the 
current operators (see \cite{Ama96}). A sum is then performed over the spin
$s'$ and the orientation $M_B$ of the residual nucleus, since in the present 
work it is assumed that the polarizations of the final particles are 
undetected. The factor $Mp'$ in eq.~(\ref{tenns}) comes from the change of 
normalization in the 
outgoing nucleon wave function, {\it i.e.,} in the CSM the particle
wave functions are normalized with a $\delta$-function containing energies, 
while the outgoing nucleon wave functions, eq.~(\ref{onda-plana}), are
standard plane waves normalized with a $\delta$-function of momentum.
The factorization of the hadronic tensor follows from the
translational invariance of the nuclear electromagnetic current,
\begin{equation}
 \label{current-spin-matrix}
\langle \np'r'|J^{\mu}(\nq)|\np r\rangle 
=\delta(\np+\nq-\np')J^{\mu}(\np',\np)_{r'r},
\end{equation}
thereby yielding the current
spin-matrix  $J^{\mu}(\np',\np)_{r'r}$ to be used below.
The most general form of this matrix is
\begin{equation}
J^{\mu}=a^{\mu}+i\nb^{\mu}\cdot\nsigma.
\end{equation}

We must make the approximation that the ground state
$|A\rangle$ has no components with momentum $\np'$, that is,
$a_{\np's'}|A\rangle\simeq 0$. As a consequence, one can easily 
show that the hadronic tensor can be written as a trace of the
product of two spin matrices
\begin{equation}
W_B^{\mu\nu}=Mp' {\rm Tr}\left[ w^{\mu\nu}(\np',\np)n_B(\np)\right],
                                       \label{h-tensor-tr}
\end{equation}
where the missing momentum $\np= \np'-\nq$ has been introduced,
the single-nucleon tensor is given by
\begin{equation}
w^{\mu\nu}(\np',\np)_{rr'} = 
    \sum_s J^{\mu}(\np',\np)^*_{sr}
           J^{\nu}(\np',\np)_{sr'},
\end{equation}
and $n_B(\np)$ is the partial momentum distribution matrix
for residual nucleus $|B\rangle$
\begin{equation}
n_B(\np)_{r'r}=
\sum_{M_B}
\langle B|a_{\np r}|A\rangle^*
\langle B|a_{\np r'}|A\rangle. \label{ea3}
\end{equation}
Note that the diagonal element $n_B(\np)_{rr}$ is the probability
that the nucleus $|A\rangle$ be a residual nucleus $|B\rangle$ plus
a particle with momentum $\np$ and third spin component $r$. 
Note also that the responses defined in eqs.~(\ref{componentes1},
\ref{componentes2}) involve combinations of the form 
\begin{equation} \label{unprimed-combinations}
w^K \sim J^{\mu\dagger}J^{\nu}+J^{\nu\dagger}J^{\mu}
= 2(a^{\mu}a^{\nu}+\nb^{\mu}\cdot\nb^{\nu}),
\end{equation}
for $K=L,T,TL$ and $TT$, while the response functions defined in
eq.~(\ref{componentes3}) involve combinations of the form 
\begin{equation} \label{primed-combinations}
w^{K'} \sim i(J^{\mu\dagger}J^{\nu}-J^{\nu\dagger}J^{\mu})
= -2(a^{\mu}\nb^{\nu}-a^{\nu}\nb^{\mu}+
     (\nb^{\mu}\times\nb^{\nu}))\cdot\nsigma,
\end{equation}
for $K'=T',TL'$. Therefore the single-nucleon responses 
$w^K$ are proportional to the unit matrix in spin space and can be written
as 
\begin{equation}
w^K=w_{S}^K,  \label{sin-nuc-res-matr-S}
\end{equation}
where the sub-index $S$ means ``scalar", while the responses $w^{K'}$
are proportional to the Pauli matrices and can be written as
\begin{equation} \label{sin-nuc-res-matr-V}
w^{K'}={\bf w}^{K'}_V\cdot\nsigma, 
\end{equation}
where the sub-index $V$ means ``vector".

\subsubsection{Momentum distribution of a shell}
Remember that our initial nuclear state 
is a polarized one-hole state,
\begin{equation}
|A\rangle=\sum_{m_i}{\cal D}_{m_ii}^{(i)}(\Omega^*)
           b^{\dagger}_{im_i}|C\rangle, \label{eqn58}
\end{equation}
while the  final states of the residual nucleus involve two holes in the core,
\begin{equation}
|B\rangle=b^{\dagger}_h b^{\dagger}_i|C\rangle = 
|h^{-1}m_h,i^{-1}m_i'\rangle. \label{eqn59}
\end{equation}
Following ref.~\cite{Gar95} we introduce the scalar $M^S$
and vector ${\bf M}^V$ momentum distributions, defined by
\begin{equation}   \label{m-distribution-matrix}
n_B(\np)= \frac12(M^S+{\bf M}^V\cdot\nsigma), 
\end{equation}
where now $M^S$ and ${\bf M}^V$ are independent of the spin indices.
In order to compute these quantities, 
 we must consider two cases, depending on the shell where the
hole $h$ is located.

In the first case we have $h\ne i$ and obtain the
unpolarized momentum distribution of a complete shell
\begin{equation} \label{62}
n_B(\np)_{rr'}=n^{(h)}_{rr'}(\np)_{\rm unpol}=
\delta_{rr'}\frac1{8\pi}[j_h]^2|\tilde{R}_h(p)|^2,
\end{equation}
where 
\begin{equation}
\tilde{R}_h(p)=\sqrt{\frac{2}{\pi}}
             \int_0^{\infty}{\rm d}r\, r^2j_{l_h}(pr)R_h(r)
\end{equation}
is the radial wave function in momentum space. In this case 
we have for the scalar and vector momentum distributions:
\begin{equation}
M^S(\np)=\frac1{4\pi}[j_h]^2|\tilde{R}_h(p)|^2,
\kern 1cm
{\bf M}^V(\np)=0.
\end{equation}

In the second case we have $h=i$. In appendix C we show
that the momentum distribution of this shell is given by the
momentum distribution of the complete (unpolarized) shell minus
the (polarized) momentum distribution of the hole:
\begin{equation} \label{65}
n_B(\np)_{r'r}=
n^{(i)}_{r'r}(\np)_{\rm unpol}-
n^{(i)}_{r'r}(\np,\Omega^*)_{\rm hole},
\end{equation}
where the scalar and vector momentum distributions for a hole
are given by
\begin{eqnarray}
M^S &=& 2\sum_{\Jb}P^+_{\Jb}f^i_{\Jb}A_{\Jb}[j_i]^2
        |\tilde{R}_i(p)|^2
        \left[  Y_{\Jb}(\Omega^*)\otimes 
                Y_{\Jb}(\hat{\np})
        \right]_{00} \label{66}\\
M^V_{\alpha}
    &=& \frac{2}{\sqrt{3}}\sum_{\Jb}\sum_{\Jb'=\Jb\pm1}
        P^-_{\Jb}f^i_{\Jb}A_{\Jb\Jb'}[j_i]^2
        |\tilde{R}_i(p)|^2
        \left[  Y_{\Jb}(\Omega^*)\otimes 
                Y_{\Jb'}(\hat{\np})
        \right]_{1\alpha} \label{67}
\end{eqnarray}
and where $\alpha=0,\pm 1$ refers to the spherical components of the vector. 
In the above equations we use the definitions:
\begin{eqnarray}
A_{\Jb}      &\equiv& \frac12(-1)^{j_i-1/2}
                 \tresj{j_i}{j_i}{\Jb}{1/2}{-1/2}{0}
                 \label{68}\\
A_{\Jb,\Jb+1} &\equiv& \frac{2\kappa_i+\Jb+1}{[\Jb]\sqrt{\Jb+1}}A_{\Jb}
                  \label{69}\\
A_{\Jb,\Jb-1} &\equiv& \frac{2\kappa_i-\Jb}{[\Jb]\sqrt{\Jb}}A_{\Jb}
                  \label{70}
\end{eqnarray}
and $\kappa_i\equiv (-1)^{j_i+l_i+1/2}(j_i+1/2)$.

\subsubsection{Expansions of the single-nucleon current and responses}

For the single-nucleon current we follow the formalism of ref.~\cite{Ama96}.
There we perform an expansion of the on-shell relativistic
electromagnetic single-nucleon current up to order 
$\eta=p/M$, but importantly do not expand on the dimensionless
variables $\kappa=q/2M$ and $\lambda=\omega/2M$ as was usually
done in the past. The expressions for the charge and
transverse current spin-matrices introduced 
in eq.~(\ref{current-spin-matrix})
are the following:
\begin{eqnarray}
\rho(\np',\np) & = & \rho_c+i\rho_{so}
                     (\cos\phi\sigma_y -\sin\phi\sigma_x)\delta  
                     \label{rho} \\
J^x(\np',\np) & = & iJ_m\sigma_y+J_c\delta\cos\phi \label{j1}\\
J^y(\np',\np) & = &-iJ_m\sigma_x+J_c\delta\sin\phi, \label{j2}
\end{eqnarray}
where the coordinate system is as in sect.~\ref{Polxsec},
$\delta=\eta\sin\theta$ is the variable used to
label the order of the relativistic correction
and $(\theta,\phi)$ is the direction of $\np$. 
The factors $\rho_c$ (charge), $\rho_{so}$ (spin-orbit),
$J_m$ (magnetization) and $J_c$ (convection) are only
$(q,\omega)$-dependent,
\begin{eqnarray}
\rho_c    &=& \frac{\kappa}{\sqrt{\tau}}G_E \\ 
\rho_{so} &=& \frac{2G_M-G_E}{\sqrt{1+\tau}}\frac{\kappa}{2}\\
J_m       &=& \sqrt{\tau}G_M \\
J_c       &=& \frac{\sqrt{\tau}}{\kappa}G_E
\end{eqnarray}
and $G_E$, $G_M$ are the electric and magnetic form factors of
the nucleon for which we use the Galster parametrization \cite{Galster}.

Using these expansions and 
eqs.~(\ref{unprimed-combinations},\ref{primed-combinations})
we obtain the following expressions for the single-nucleon responses:
\begin{eqnarray}
w^L    & = & \rho_c^2+\rho_{so}^2\delta^2 \label{snL}
             \label{SNR-L}\\
w^T    & = & 2J_m^2+J_c^2\delta^2 
             \label{SNR-T}\\
w^{TL} & = & 2\sqrt{2}(\rho_cJ_c+\rho_{so}J_m)\delta\cos\phi 
             \label{SNR-TL} \label{eqtl}\\
w^{TT} & = & -J_c^2\delta^2\cos2\phi 
             \label{SNR-TT} \label{eqtt}\\ 
w^{TL'} & = & 2\sqrt{2}   \left[\rho_cJ_m\nne_1+
       J_c\rho_{so}\delta^2\sin\phi(\cos\phi\,\nne_2-\sin\phi\,\nne_1)
       \right. \nonumber\\
        &   & \left.\mbox{}
         -\rho_{so}J_m\delta\cos\phi\,\nne_3
       \right] \cdot\nsigma
               \label{SNR-TL'}\\
w^{T'}  & = & 2\left[J_cJ_m\delta(\cos\phi\,\nne_1+\sin\phi\,\nne_2)
                     -J_m^2\nne_3
               \right]\cdot\nsigma \label{snT'}
               \label{SNR-T'}
\end{eqnarray}
As already noted, the $w^K$ responses are spin independent, and
all of the spin dependence goes into the $w^{K'}$ responses.

\subsubsection{Inclusive nuclear responses}

From eqs.~(\ref{h-tensor-tr}), (\ref{sin-nuc-res-matr-S}),
(\ref{sin-nuc-res-matr-V}) and (\ref{m-distribution-matrix}) 
 we can write 
the exclusive responses for a shell as
\begin{eqnarray}
{\cal R}^K 
& = & Mp'\, w^K_SM^S(\np)\\
{\cal R}^{K'} 
& = & Mp'\,\nw^{K'}_V\cdot\nM^V(\np).
\end{eqnarray}
We obtain the inclusive responses for given proton and neutron shells 
by integration over the 
angles $\Omega'=(\theta',\phi')$ of the outgoing nucleon. As the momentum
density for a shell is a function of $\np=\np'-\nq$,
it is convenient to perform the integral using the variables
$(p,\phi)$ instead of $\Omega'$. Note that $\phi'=\phi$ because
we have chosen the $z$-axis along $\nq$ and that from
$p^2=p'^2+q^2-2p'q\cos\theta'$ we obtain ${\rm d}\cos\theta'=-p{\rm d}p/p'q$.
Accordingly the inclusive responses of a shell can be written
\begin{eqnarray}
R^K &=& \frac{M}{q}
        \int_{|p'-q|}^{p'+q}{\rm d}p\,p
        \int_0^{2\pi}{\rm d}\phi\,\,
        w^K_SM^S(\np) 
        \label{90a}\\
R^{K'}&=& \frac{M}{q}
           \int_{|p'-q|}^{p'+q}{\rm d}p\,p
           \int_0^{2\pi}{\rm d}\phi\,\,
           \nw^{K'}_V\cdot\nM^V(\np)
           \label{90b}
\end{eqnarray}
For a complete (unpolarized) shell we have only the scalar momentum
distribution; hence $\nM^V=0$ and the $T'$ and $TL'$ responses are zero. 
From eq.~(\ref{62}) we see that  $M^S(\np)$ depends only  on $p$.
The single-nucleon responses TL and TT in eqs.~(\ref{eqtl},\ref{eqtt}) are 
proportional to $\cos\phi$ and $\cos2\phi$, respectively, and therefore their
contributions go away upon integration on  $\phi$ --- 
as expected, a complete shell only has two inclusive responses
$L$ and $T$. In particular, for the expansions in 
eqs.~(\ref{rho}-\ref{j2}) the only nonzero reduced responses 
are given by:
\begin{eqnarray}
W_0^L &=&\frac{[j_i]}{4\pi}R^L=
         \frac{[j_i]}{2}
        \frac{M}{q}\int_{|p'-q|}^{p'+q}{\rm d}p\, p
        (\rho_c^2+\rho_{so}^2\delta^2)M^S(p) 
         \label{complete-shell-L}\\
W_0^T &=&\frac{[j_i]}{4\pi}R^T=
         \frac{[j_i]}{2}
         \frac{M}{q}\int_{|p'-q|}^{p'+q}{\rm d}p\, p
        (2J_m^2+J_c^2\delta^2)M^S(p).
         \label{complete-shell-T}
\end{eqnarray}

From eq.~(\ref{65}) we see that the responses of a shell with a hole 
can be obtained as the response of the complete shell (given
above) minus the response of a single hole.  
In appendix C we show that the reduced inclusive responses for a
hole are given by
\begin{equation} \label{94}
\left( {W}^{K,K'}_{\Jb}\right)_{\rm hole}
                =-\frac{M}{q}\int_{|p'-q|}^{p'+q}{\rm d}p\, p
                \frac{[j_i]}{4\pi}|\tilde{R}_i(p)|^2
                \Phi_{\Jb}^{K,K'}(q,\omega,\cos\theta)
                \label{red-res-hol-PWIA},
\end{equation}
where the functions $\Phi$  depend on $p$ through the function
\begin{equation}
\cos\theta=(p'{}^2-p^2-q^2)/2pq \label{costh}
\end{equation}
and they have the following forms to order $\delta^2$:
\begin{eqnarray}
\Phi_{\Jb}^L    &=& (\rho_c^2+\rho_{so}^2\delta^2)
                 [\Jb]A_{\Jb}P^0_{\Jb}(\cos\theta)
                 \label{phi-L}\\ 
\Phi_{\Jb}^T    &=& (2J_m^2+J_c^2\delta^2) 
                 [\Jb]A_{\Jb}P^0_{\Jb}(\cos\theta)
                 \label{phi-T}\\ 
\Phi_{\Jb}^{TL} &=& (\rho_cJ_c+\rho_{so}J_m)\delta
                 \frac{2\sqrt{2}[\Jb]}{\Jb(\Jb+1)}
                 A_{\Jb}P^1_{\Jb}(\cos\theta)
                 \label{phi-TL}  \\ 
\Phi_{\Jb}^{TT} &=& -J_c^2\delta^2 
                 \frac{[\Jb]}{(\Jb-1)_4}
                 A_{\Jb}P^2_{\Jb}(\cos\theta)
                 \label{phi-TT}\\ 
\Phi_{\Jb}^{T'} &=& 2J_m^2 
                 \sum_{\Jb'=\Jb\pm1} 
                 [\Jb][\Jb']
                 A_{\Jb\Jb'}
                 \tresj{\Jb}{\Jb'}{1}{0}{0}{0}
                 P^0_{\Jb'}(\cos\theta)\nonumber\\ 
             & & \mbox{}+
                 J_cJ_m\delta
                 \sum_{\Jb'=\Jb\pm1} 
                 \frac{2\sqrt{2}[\Jb][\Jb']}{\sqrt{\Jb'(\Jb'+1)}}
                 A_{\Jb\Jb'}
                 \tresj{\Jb}{\Jb'}{1}{0}{1}{-1}
                 P^1_{\Jb'}(\cos\theta)
                 \label{phi-T'}\\ 
\Phi_{\Jb}^{TL'}&=&\rho_cJ_m 
                 \sum_{\Jb'=\Jb\pm1} 
                 \frac{4[\Jb][\Jb']}{\sqrt{\Jb(\Jb+1)}}
                 A_{\Jb\Jb'}
                 \tresj{\Jb}{\Jb'}{1}{1}{0}{-1}
                 P^0_{\Jb'}(\cos\theta)\nonumber\\
            & & -\rho_{so}J_m\delta
                 \sum_{\Jb'=\Jb\pm1} 
                 \frac{2\sqrt{2}[\Jb][\Jb']A_{\Jb\Jb'}}%
                      {\sqrt{\Jb(\Jb+1)\Jb'(\Jb'+1)}}
                 \tresj{\Jb}{\Jb'}{1}{1}{-1}{0}
                 P^1_{\Jb'}(\cos\theta)\kern 1cm\nonumber\\
            & & -\rho_{so}J_c\delta^2
                 \sum_{\Jb'=\Jb\pm1} 
                 \frac{2\sqrt{2}[\Jb][\Jb']A_{\Jb\Jb'}}%
                      {\sqrt{\Jb(\Jb+1)(\Jb'-1)_4}}
                 \tresj{\Jb}{\Jb'}{1}{-1}{2}{-1}
                 P^2_{\Jb'}(\cos\theta)\nonumber\\
            & & -\rho_{so}J_c\delta^2
                 \sum_{\Jb'=\Jb\pm1} 
                 \frac{2\sqrt{2}[\Jb][\Jb']A_{\Jb\Jb'}}%
                      {\sqrt{\Jb(\Jb+1)}}
                 \tresj{\Jb}{\Jb'}{1}{1}{0}{-1}
                 P^0_{\Jb'}(\cos\theta),
                 \label{phi-TL'} 
\end{eqnarray}
with $(\Jb-1)_4\equiv (\Jb-1)\Jb(\Jb+1)(\Jb+2)$.
These equations are very useful to get a feeling about the size
of the first- and second-order relativistic corrections in the
responses, in contrast to the situation that occurs in the CSM where it is 
more difficult to know {\it a priori} the actual importance of the various 
orders. Note that in particular in the complete shell model
calculation the contributions to order $\delta$ and $\delta^2$ 
cannot be independently separated in general. 
Moreover, it is also useful to compare with the results
using the complete relativistic expressions for the 
single-nucleon responses in the factorized approximation. It is also
very instructive, as we shall see in the next section, to
compare these results with the ones obtained in the CSM
where the same expansion for the single-nucleon
current operator is used.

The terms proportional to $\delta^n$ in eqs.~(\ref{phi-L}-\ref{phi-TL'})
will be of order $\eta_F^n$ after integration over $p$, where 
$p_F$ is some average nucleon momentum for the shell $i$ (of order the 
Fermi momentum which implies that typically 
$\eta_F$ will be of the order $1/4$). 

From eqs.~(\ref{complete-shell-L},\ref{complete-shell-T}) we see
that in PWIA there are no first-order terms in $\eta_F$ 
(there is no interference between charge and spin-orbit or
between magnetization and convection pieces).
Therefore the first relativistic correction to the static 
(longitudinal) charge and (transverse) spin responses is of
$O(\eta_F^2)$, that is, very small as we shall see. 
The same considerations are applicable to the $L$ and $T$ responses for a hole. 

As for the $TL$ response, it is of $O(\eta_F)$ and we see
that the spin-orbit and convection contributions enter at the same level.
Thus, at high $q$ we expect that the effect of the
spin-orbit term in this response (not included usually in non-relativistic
calculations) has the same degree of importance as the convection current.

The $TT$ response is of $O(\eta_F^2)$, explaining why it is always very small
for quasielastic scattering. Strictly speaking we should neglect this 
contribution not only because it is small, but also because our expansion of 
the relativistic four-current has been truncated at $O(\eta)$. That is, to be 
consistent in computing the responses we should include only terms up to 
$O(\eta)$. However, it is useful to evaluate the $O(\eta^2)$ effects even 
if incompletely to see whether or not they could have a significant impact 
on the $TT$ response. Indeed, we shall see in the next section that this 
response is very sensitive to the distortion of the outgoing nucleon and 
therefore to details of the nuclear model. 

The $T'$ and $TL'$ responses are of $O(\eta_F^0)$ and so are
expected to be larger than the TT and TL responses.
The first-order  correction to these responses is due to
the spin-orbit piece in the $TL'$ response 
and to the convection current contribution in the $T'$ response.

\section{Results}\label{Results}

\subsection{Comparisons of CSM and PWIA}\label{Comparisons}
Before showing results for more realistic nuclei, and in a first test 
of our formalism, it is useful to compare our model with an exactly
soluble problem, namely, the simpler process of elastic scattering from a 
polarized proton at rest. Obviously, in that case the cross section
is zero unless $\omega=-Q^2/2M$, implying that a factor
$\delta(\omega+Q^2/2M)$ appears in the responses. Aside from that factor, 
the polarized responses in this case are given by \cite{Don86}:
\begin{eqnarray}
{W}^L_0     & = & (1+\tau)G_E^2 \\
{W}^T_0     & = & 2\tau G_M^2 \\
{W}^{TL'}_1 & = & 2\sqrt{2}\sqrt{\tau(1+\tau)}G_MG_E \\
{W}^{T'}_1  & = & -2\tau G_M^2.
\end{eqnarray}
Here the initial spin is $j_i=1/2$ and we
have only four responses corresponding to $\Jb=0,1$. Two are equal in 
magnitude, ${W}^{T'}_1 = -{W}^{T}_0$, 
and inserting the typical value $q=500$ MeV/c we see that we have
${W}^{TL'}_1  \simeq  2{W}^{L}_0$. 

Now we compare these results with an extreme nuclear model that has 
one proton in the $1s_{1/2}$ shell 
(actually we show the results for the well parameters of $^{12}$C,
although the specific nucleus is not essential for the test). 
In this case we expect that the above elastic scattering results will be 
broadened in energy due to the momentum distribution of the proton,
but that the integral over $\omega$ of these responses approximately 
reproduces the elastic case. In particular, the relative order of magnitude 
of the different elastic responses seen above should be preserved, that is, 
the $TL'$ response should be about two times the $L$ response, while
the $T'$ responses should be about minus the $T$ response.
We can see in fig.~1 that these expectations are approximately
met. Therein we show the different responses computed in the CSM 
and PWIA (on-shell) in the solid and dashed curves, respectively. 
Although our purpose in this initial discussion is only to get a feeling 
for the qualitative behaviour of the responses, at this juncture we 
note already that the CSM and PWIA (on-shell) models give similar results, 
the most obvious difference being the clear energy shift that we discuss 
below. All of the responses seem to differ only by a scale factor, as do 
their integrals. The value of the $L$ and $TL'$ responses  at the
maximum are, respectively,  about 0.3 and 0.6~GeV$^{-1}$, 
while the $T$ and $T'$ ones take the values 0.3 and
$-0.3$~GeV$^{-1}$, respectively. 

Thus, the structure for all of the polarized responses for a proton in the 
$1s_{1/2}$ shell appears to be very easily accounted for, in contrast with the 
situation seen for nucleons in other shells. In particular, we display 
results in fig.~2 for a proton in the $1p_{1/2}$ or $2s_{1/2}$ shell, the 
one-hole nuclei $^{15}$N and $^{31}$P with well parameters taken from 
$^{16}$O and $^{40}$Ca, respectively, again total angular momentum 1/2 
situations. Note that we do not really expect that the simple one-hole 
model should be very successful for mid-shell cases such as $^{31}$P; it is 
used here mainly for illustrative purposes, although, as discussed below, 
polarized inclusive quasielastic electron scattering may provide a 
very powerful probe of configuration-mixing effects such as those 
expected for cases such as $^{31}$P. The $1p_{1/2}$ and $2s_{1/2}$ 
polarized responses obtained at a typical value of the 
momentum transfer $q=500$ MeV/c are shown in fig.~2. Here and in several of 
the cases to follow we omit the unpolarized responses ${W}_0^{L,T}$; since 
these involve all of the nucleons, and not just the single valence particle 
carrying the nucleus' angular momentum in our extreme one-hole model, they 
behave rather similarly for all nuclei (compare, for instance, figs.~1 and 5 
given below). Their magnitudes, of course, depend 
on factors that involve the charge $Z$ and neutron number $N$ of the given 
nucleus and their widths increase with Fermi momentum $p_F$. The polarized 
responses for the $1s_{1/2}$ and $2s_{1/2}$ cases, while differing in 
detail because they have different momentum distributions, are similar in 
magnitude and occur with the same signs. In contrast, the $1p_{1/2}$ cases 
are rather different: first, their signs are usually the opposite from the 
$s$-wave cases and, second, the energy distributions now reflect the 
$p$-wave nature of the momentum distribution. The signs in the $p_{1/2}$ 
case are easily seen to arise from the fact that the total angular 
momentum is comprised of one unit of orbital angular momentum coupled with 
spin-1/2 in a ``jack-knifed'' configuration in which spin and total angular 
momentum projections oppose --- said another way, if the total angular 
momentum is polarized in some given direction then the spin points in 
the same (opposite) direction for $s_{1/2}$ ($p_{1/2}$) states.

The PWIA (on-shell) formalism presented in the last section is very 
useful as it provides some insight into the behaviours observed for these 
special spin-1/2 cases. Starting with the $s_{1/2}$ case, by using 
eqs.~(\ref{phi-T'},\ref{phi-TL'}) for the 
leading-order dependences in the integrals in eq.~(\ref{red-res-hol-PWIA}) that 
determine the reduced responses in this model, 
\begin{equation}
\Phi_1^{T'}=\frac{1}{\sqrt{2}} J_m^2, \qquad\qquad \Phi_1^{TL'}=-\rho_c J_m,
\end{equation}
we see that the integrands have no $\theta$-dependence. 
This is the same situation that occurs for the integrals 
in eqs.~(\ref{complete-shell-L},\ref{complete-shell-T}) for
all unpolarized responses. These integrals are then simply of the form
\begin{equation}
W\propto \int_{|y|}^{2q+y} pdp \rho (p),
\end{equation}
where $\rho(p)$ is the appropiate momentum distribution ($M^S (p)$ in 
the unpolarized cases and $-|{\tilde R}_i (p)|^2$ for the $s_{1/2}$ polarized 
cases; note that for these one-hole cases the minus sign of eq.~(\ref{65}) 
enters for the polarized responses). Here we have used an approximation for 
the limits of integration by neglecting binding effects --- these are, of 
course, included in the actual results presented here --- whereby
$p'\cong \sqrt{2M\omega (1+\omega/2M)}$, implying that $p'-q\cong y$ 
with $y$ the usual scaling variable (see ref.~\cite{Don84} and also several 
approximations discussed in ref.~\cite{Cen96}). In the scaling limit where 
$q\rightarrow\infty$ the upper limit may safely be taken to $\infty$, 
{\it i.e.,} larger than any characteristic nuclear momentum; on the other 
hand, the lower limit $|y|$ determines how much of the relevant momentum 
distribution is integrated. When $y=0$ the full range of integration is 
covered and, as usual, the unpolarized responses attain their maximum 
values, defining the quasielastic peak. In the cases of the $2s$ polarized 
responses the situation is similar, except that only the valence (polarized) 
nucleon's momentum distribution enters. Starting from very low 
$\omega$ ($y<0$) little 
of the momentum distribution is covered and the response is small. As 
$\omega$ increases, the response also increases until the higher-$p$ part 
of $|{\tilde R}_{2s}|^2$ is covered, ``stalls'' as the momentum 
integration passes through the node in the wave function ($p\cong 120$ MeV/c 
which corresponds to $\omega\cong 75$ and 185 MeV) and then 
increases again as the lower-$p$ part of the momentum distribution is 
covered. The quasielastic peak occurs as in the unpolarized situation 
near $y=0$ and then the pattern repeats as $y$ continues to rise beyond zero.
Consequently the $2s$ polarized responses have the behaviour seen in 
fig.~2 with ``shoulders'' on the sides of the quasielastic peak. The fact 
that the response is not symmetrical reflects the behaviour of the 
single-nucleon form factors which vary with $\omega$ through their 
dependences on $|Q^2|=q^2-\omega^2$.

The polarized responses in the $p_{1/2}$ case are a little more 
complicated in that the integrands in eq.~(\ref{red-res-hol-PWIA}) contain
\begin{equation}
\Phi_1^{T'}=\frac{1}{\sqrt{2}} J_m^2 \cos 2\theta, 
  \qquad \Phi_1^{TL'}=\rho_c J_m \cos^2 \theta
\end{equation}
which have $\theta$-dependences through factors of $\cos 2\theta$ and 
$\cos^2 \theta$, these in turn being functions of $p$, $q$ and $y$ 
(see eq.~(\ref{costh})) through
\begin{equation}
\cos\theta ={\frac{1}{2pq}}(2qy+y^2-p^2),
\end{equation}
again neglecting binding effects for this argument.
When $p=|y|$, one has $\cos\theta=+1(-1)$ for $y>0(y<0)$ and as $p$ becomes 
larger $|\cos\theta|$ falls off with characteristic half-width of $2y$. Thus, 
for very small $y$ (near the quasielastic peak) $|\cos\theta|$ is only large
for a small region where $p$ is close to $|y|$. This implies that the $TL'$ 
response whose integral contains the factor $\cos^2 \theta$ should be 
small, while the $T'$ response, involving $\cos 2\theta=2\cos^2 \theta -1$, 
should be large in magnitude, as observed in fig.~2. As $|y|$ increases 
away from zero, the sampled region widens. In the $TL'$ response the 
characteristic ``double-bump'' behaviour reflects the $1p$ momentum 
distribution where the peak positions are determined by the maximum in that 
distribution, namely about 100 MeV/c; in the $T'$ response the factor 
$\cos 2\theta$ changes sign as the integral progresses from $p=|y|$ to 
higher values and accordingly can also produce a sign-change in the response 
itself.

Once the total angular momentum is 3/2 or larger the situation is clearly 
more complicated. Specifically, for $j_i=3/2$ one now has ${\cal J}=2$ and 3 
(tensor- and octupole-polarized) responses in addition to ${\cal J}=0$ 
(unpolarized) and 1 (vector-polarized) responses. For a $1p_{3/2}$ proton, 
corresponding to the one-hole nucleus $^{11}$B and 
using well parameters for $^{12}$C, we obtain the results shown 
in figs.~3 and 4 again for $q=500$ MeV/c (omitting the unpolarized responses 
for the reasons stated above). We observe first of all that in this 
``stretched'' configuration (spin and orbital angular momentum projections 
in the same direction) the vector-polarized responses are again 
qualitatively the same as the $s$-wave cases discussed above. The tensor- 
and octupole-polarized responses, on the other hand, have no analogs in the 
spin-1/2 situations. These range in magnitude from being comparable to 
the vector-polarized responses (${W}_3^{T'}$) to being a factor of 
about two smaller (${W}_2^{L}$, ${W}_2^{T}$, ${W}_3^{TL'}$) 
to being more than an order of magnitude smaller (${W}_2^{TL}$, 
${W}_2^{TT}$). The corresponding $1d_{3/2}$ ``jack-knifed'' results, 
{\it i.e.,} for the case of $^{39}$K, are shown in figs.~5 and 6 
(all of the responses are now given, since we shall return to discuss 
the results for this nucleus in more detail in the next two subsections). 
Here again the vector-polarized responses have the sign they do for the 
other ``jack-knifed'' case considered above, namely, $1p_{1/2}$. The 
magnitudes of the $1d_{3/2}$ and $1p_{3/2}$ are comparable, although, 
given the degree of structure observed, it is difficult to relate the 
results to any simple picture of the state occupied by the polarized nucleon.

The reason why the polarized unprimed responses are similar but the primed
are so different for equal $j_i$ and different $l_i$ can be explained in PWIA. 
In fact, we see from eqs.~(\ref{red-res-hol-PWIA},\ref{phi-L}-\ref{phi-TL'}) 
that the polarized unprimed responses depend on the orbital
angular momentum $l_i$ of the hole only via the momentum distribution
$|{\tilde R}_i(p)|^2$ of the complete shell $i$. These distributions 
must satisfy the sum rule $\int |{\tilde R}_i(p)|^2=1$ and thus 
we can suppose, as a first approximation that the integrals extend 
between the limits $|p'-q|$ and $p'+q$ for high $q$ and are not very
different for $l_i=j_i\pm 1/2$, so the quasielastic responses do not
depend critically on $l_i$. Where the primed responses are concerned, 
we can see from eqs.~(\ref{phi-T'},\ref{phi-TL'})
that they also depend on $l_i$ via the coefficients 
$A_{\Jb\Jb'}$ given in eqs.~(\ref{69},\ref{70}). As a consequence, the 
primed responses are very sensitive to the angular momentum $l_i$
of the polarized shell.

Thus we see quite dramatically different behaviour when the quantum numbers 
of the nucleon carrying the polarization are varied. In a configuration-mixed 
situation such as occurs in the middle of the $1p$-shell we would expect 
that the polarization effects would arise from a combination of $1p_{3/2}$ 
and $1p_{1/2}$ contributions. Since the extreme, pure one-hole results 
presented here are so different for these two configurations, even changing 
sign, it is clear that the polarized responses have the potential to 
provide a sort of ``configuration analyzer''. Another example is the $^{31}$P 
case introduced above, where we do not expect the pure $2s_{1/2}$ proton hole 
model to be especially good and instead expect that configuration-mixing in 
the $2s-1d$ shell should be important. By examining, for instance, the 
vector polarized responses shown in the figures, it is again clear that 
the sensitivity to details of the configuration-mixing is high and that 
any significant deviation from the extreme $2s_{1/2}$ results given in fig.~2 
and $1d_{3/2}$ results given in fig.~6 for ${W}^{T',TL'}_1$ should be 
observable.

Let us now return to compare in more detail the results obtained for the 
two basic models considered in this work, the CSM and PWIA(on-shell) models; 
we return to make further connections with the PWIA(CC1) model at the 
beginning of the next subsection. We expect that the CSM predictions should be 
more representative of what will be measured when experiments are performed on 
polarized nuclei of the type we are considering in this work when the 
outgoing nucleon's energy is relatively small, {\it i.e.,} when the momentum 
transfer and hence the energy transfer at the quasielastic peak 
are relatively small. On the other hand, once the kinematics progress to 
the higher-energy regime, the FSI are expected to play a more minor role and 
the PWIA should be expected to become valid. We may take as ``typical'' 
regimes for the two situations conditions in which the outgoing nucleon is 
below say 50 MeV, where the FSI are surely quite strong and where (at 
least) the distortion effects included in the CSM treatment should be 
necessary, and at or above 200 MeV, where the real central nucleon-nucleus 
potential is known from optical model analyses of nucleon scattering to be 
rather small compared with its value at low or negative energies and thus 
where the PWIA should be adequate. From the relative simplicity of the 
PWIA formalism presented in this work it is clear that the latter when 
applicable has the advantage of providing a more direct connection to 
the desired initial-state nuclear structure issues. Part of our motivation 
in the present work is to assess the degree of sensitivity to FSI effects 
one should expect to see in the various inclusive polarized response 
functions through the use of the two extreme models.

Where the FSI distortion effects are concerned we have already seen above 
that the general behaviour of the responses in the CSM 
and PWIA models is similar (with the exception of the $TT$ response that is 
discussed in more detail below). In particular, the Coulomb sum rule (the 
zeroth energy-weighted moment of ${W}^L_0$ reduced in the standard way; 
see ref.~\cite{Cen96}) is virtually the same and equal to unity in the two 
models for high momentum transfers. The plane-wave results are, on the average, 
shifted to the right with respect to the CSM responses. By forming the 
first energy-weighted sum rule \cite{Cen96} one sees that this shift is 
more or less constant as a function of $q$, whereas the width of the 
reduced response (the variance, see ref.~\cite{Cen96}) is rather close 
to the same value in the two models.

The shift may be understood using the following argument: in the CSM the 
energy of the ejected nucleon $\epsilon_p=t_p+v_p$ is the sum
of kinetic plus potential energy, while in PWIA the energy
of the nucleon $\epsilon'_p=t'_p$ is only kinetic.
As the two energies are the same, $\epsilon_p=\epsilon_p'$,
we have $t'_p<t_p$ because the potential energy is negative.
This means that the velocity of the nucleon knocked-out
from the nuclear interior is greater for distorted than for plane
waves, implying a shift of the responses. An alternative way of seeing this 
is the following: let us suppose that at high energy  the matrix elements 
of the current computed in the PWIA or CSM are approximately the same, 
$\langle J^{\mu} \rangle_{PW}\simeq \langle J^{\mu} \rangle_{CSM}$, 
that the potential energy for the particle 
is a constant $v_p\simeq -V<0$, and write $t_p=\epsilon_p+V$. Then we
can write the PWIA hadronic tensor schematically as
\begin{eqnarray}
W^{\mu\nu}_{PW}(q,\omega)
& = & \sum\delta(t_p-\epsilon_h-\omega) 
      \langle J^{\mu} \rangle_{PW}^*
      \langle J^{\nu} \rangle_{PW} \nonumber\\
& \simeq & \sum\delta(\epsilon_p+V-\epsilon_h-\omega) 
      \langle J^{\mu} \rangle_{CSM}^*
      \langle J^{\nu} \rangle_{CSM} \nonumber\\
& = & W^{\mu\nu}_{CSM}(q,\omega-V).
\end{eqnarray}
This means that the PWIA responses will be shifted to the right  
of the CSM ones by an amount $V$ which should arise as the average of 
the Woods-Saxon potential, in our case about  $\sim
30$--$35$ MeV, namely, approximately the order of the observed shifts.

Let us now return to a more detailed treatment of the 
case of $^{39}$K treated as a $1d_{3/2}$ proton hole in closed shell 
$^{40}$Ca (for treatment of exclusive electron scattering from polarized
$^{39}$K and $^7$Li see ref. \cite{Bof88}).
In figs.~5 and 6 results were shown for $q=500$ MeV/c; now these are 
extended both to lower (300 MeV/c in figs.~7 and 8) and higher 
(700 MeV/c in figs.~9 and 10) momentum transfers.
The naive quasielastic peak occurs at $\omega=|Q^2|/2M=\sqrt{q^2+M^2}-M$, 
namely at $\omega=47$, 125 and 232 MeV for $q=300$, 500 and 700 MeV/c, 
respectively. Our expectation from the above arguments is thus that at 
300 MeV/c the CSM should be more in accord with reality, since FSI should be
strong at such low energies. In contrast, at 700 MeV/c at the quasielastic 
peak the outgoing nucleon is energetic enough that the FSI should be weak 
and the PWIA, not the CSM (which has the same potential acting in the 
final and initial states), should be more in line with the actual dynamics.

The results in figs.~5--10 show that, aside from the shift, the CSM and 
PWIA responses become rather similar at high momentum transfer, the 
exceptional case being the $TT$ response. For low momentum transfer 
the basic structures still persist, although now some notable differences 
beyond a simple shift are observed; see for instance the ${W}^{TL'}_1$, 
${W}^{T'}_3$ and ${W}^{TL'}_3$ responses in fig.~8.

Now we discuss the $TT$ response. This is the only case
where the CSM and PWIA yield very different results for the structure 
and magnitude of the responses. The reason is the following:
on the one hand, in PWIA the zeroth- and first-order terms in the expansion 
are exactly zero, due to cancelations in the spin sums, and consequently
only the very small term involving the convection current survives.
On the other hand, the terms that were exactly zero in PWIA are not zero 
for the CSM, because the radial wave function for an ejected nucleon with 
$j=l+1/2$ is slightly different from the $j=l-1/2$ one, due to the spin-orbit 
term in the Woods-Saxon potential, and thus the magnetization current gives some
nonzero contribution; being of zeroth-order in $\delta$, it is enhanced with 
respect to the second-order contribution. As a consequence, the $TT$ response 
could be interesting to exploit in studying distortion and shell effects, 
since in PWIA this response would be practically zero, but the FSI enhances 
the response by more than an order of magnitude at the maxima. Unfortunately, 
even with the distortion effects present as in the CSM this response is 
rather small and would likely be a challenge to measure. 

The same argument could be applied to the $TL$ responses, since
in PWIA they have no zeroth-order contribution, but are of order
$O(\delta)$. In this case it appears that the zeroth-order 
CSM contribution is the one responsible for the differences 
between the $TL$ responses in PWIA and CSM, although in this case the 
effect is less pronounced; however, due to the larger values of these 
responses when compared with the $TT$ cases, they might be better suited 
to measurement.

\subsection{Relativistic corrections}

In this subsection we briefly discuss some of the relativistic-order 
effects treated in the present work. We begin with a comparison of 
off-shell effects by comparing in fig.~11 the $^{39}$K even-rank 
responses at $q=500$ MeV/c for the PWIA (on-shell) --- used in all other 
PWIA results given in this paper --- and the PWIA (CC1), together with the 
CSM results again for reference. Specifically, for the off-shell model 
we use the $CC1^{(0)}$ prescription discussed in ref.~\cite{Cab93}. We see 
that typically the two PWIA models 
differ by a few and even up to about 10\%, the exception again being the 
$TT$ response which is very small in PWIA but where the on-/off-shell 
behaviour is most dramatic. In exploring the $q$-dependence of these 
differences (not shown) we find very little variation in going from 
300 to 1000 MeV/c. Since we cannot know the correct form of the 
single-nucleon current, and must therefore take either model as being 
acceptable until off-shell effects can be better understood, these 
differences should be regarded as the scale of theoretical uncertainty 
stemming from nucleon model dependences --- in fact, the uncertainties 
are representative only, since other models for off-shellness could 
yield still different results. In future work we intend to examine such 
behaviour in more depth.

The second aspect we want to comment upon refers to the expansion
of the single-nucleon responses in powers of 
$\delta=\eta\sin\theta$. In the inclusive PWIA (on-shell) responses we have
maintained these terms in order to be consistent with the shell
model calculation, although clearly our treatment of second-order terms 
is incomplete since we have retained only contributions in the amplitudes of 
$O(\delta)$ and then squared them, whereas other $O(\delta^2)$ terms occur 
in the amplitudes themselves which can interfere with contibutions of 
leading-order in another amplitude. We must not find large effects from 
this inconsistency or else we would have to question the entire procedure 
and indeed when comparing results with and without the second-order terms 
we find at most a few percent difference (excepting again the very small 
$TT$ case). This is even smaller than the on-/off-shell differences 
discussed above and, at least for inclusive quasielastic electron 
scattering if not necessarily elsewhere, can be safely ignored.

Finally, it is common in some non-relativistic quasielastic scattering 
calculations to omit the convection and (or) spin-orbit terms in the 
electromagnetic current. This is a good approximation for the unpolarized 
responses, but in the polarized case it is not so obvious that they can be 
ignored since there interferences occur that can enhance one or both of the 
terms. Accordingly we end our studies here with a discussion of the effects 
that are first-order in $O(\delta)$, focusing again on the case of 
$^{39}$K in the CSM. In figs.~12 and 13 we show results at $q=500$ MeV/c 
for the total current (solid curves), for the case where only the 
zeroth-order charge and magnetization terms have been retained (dashed curves) 
and for the case where charge, magnetization and convection terms are included 
(dot-dashed curves); only the total includes the spin-orbit contributions.

In PWIA the $L$ and $T$ responses have no first-order terms, and the only 
correction is of second-order. In contrast, in the CSM the first-order terms, 
while small, are not exactly zero. In the unpolarized $L$ response 
the effect of the spin-orbit contribution is seen to be about 2.5\% and grows 
to about 3\% at 700 MeV/c and 4\% at 1000 MeV/c. In the unpolarized $T$ 
response the convection 
current produces less than 1.5\% at 500 MeV/c, falls to below 1\%
at 700 MeV/c and continues to decrease with $q$, becoming 
entirely negligible at 1000 MeV/c. In both cases the other uncertainties 
are at least comparable, as discussed above.

This behaviour should be contrasted with the effects seen in the 
interference responses. In particular, as seen in eq.~(\ref{phi-TL}), 
the $TL$ response is especially good when searching for first-order effects,
since it is of first-order even in PWIA. Again, in contrast to the PWIA 
for the CSM the zeroth-order effects are nonzero, as seen in the dashed lines 
in fig.~12, although it is clear that first-order effects are also essential. 
By combining the results shown in fig.~12 for $q=500$ MeV/c with those shown 
in fig.~14 for $q=300$ and 700 MeV/c it is clear that the relative 
importance of these effects does not go away with increasing momentum 
transfer, although the overall importance of this interference response 
function is not great and it may be difficult to isolate experimentally.
In particular, we note in comparing the dot-dashed and solid curves that the 
effect of the spin-orbit terms, being 
proportional to the momentum transfer, becomes bigger at high $q$.

With regard to the (small) $TT$ response, we see that zeroth-order
effects are dominant while in PWIA they are zero, as discussed above.
It is important to stress the fact that the dominance of zeroth-order
effects arises exclusively from the distortion of the ejected nucleon. 

Finally, from the eqs.~(\ref{phi-T'},\ref{phi-TL'}) for the $T'$ and 
$TL'$ responses in PWIA we see that in fact here there are first-order terms 
that could be important. As seen in fig.~13 the $T'$ response has 
relatively large convection current contributions that, as with the 
unpolarized $T$ response discussed above, are large at low $q$ and fall 
with increasing momentum transfer although still remaining non-negligible 
even at 1000 MeV/c. The $TL'$ response, on the other hand is harder to 
analyze: for the PWIA the spin-orbit terms contribute in first order, 
whereas for the CSM both the spin-orbit and convection current terms can 
play a role. Again the effects seen here in the figure for 500 MeV/c are 
typical of what is found even at 1000 MeV/c.

\section{Summary and conclusions}\label{Concl}

In this paper we have studied {\em inclusive} quasielastic scattering of 
polarized electrons 
from polarized nuclei, focusing on the six classes of response functions 
that occur, $L$, $T$, $TL$, $TT$ for situations where only the target is 
polarized and $T'$ and $TL'$ for situations having both polarized electrons 
and target nuclei. To illustrate the concepts involved we have obtained 
results in the extreme shell model for the case of one-hole nuclei 
employing the continuum shell model, on the one hand, and the plane-wave 
impulse approximation, on the other. Our expectation is that at moderate 
momentum transfers, where the quasielastic peak lies at low energy and 
thus where the ejected nucleons are also low-energy, the CSM with the 
same potential in the initial and final states should fairly represent the 
mean-field interaction effects, whereas the PWIA, having no FSI will 
miss these effects insofar as the final state is concerned. In contrast, at 
high momentum transfers, where the quasielastic peak and outgoing nucleon 
energies are high, the FSI are expected to be rather weak and the PWIA 
should be applicable, whereas the CSM with the full potential (which is 
used in the present work to maintain orthogonality between initial and 
final single-particle wave functions) should exaggerate such effects.

Only one-body electromagnetic 
current operators are discussed in the present work, as these are expected 
to dominate in the quasifree regime considered, although in discussing the 
special interferences that occur in such polarization studies 
more work is needed to explore other effects that 
have been found to be small for unpolarized quasielastic scattering from medium 
nuclei, such as two-body meson-exchange currents \cite{Ama94}. In the 
polarization case such effects might be enhanced by interference with the
one-body current, as happens for elastic scattering and electroexcitation 
of discrete states \cite{Ama94b}. 
MEC effects have been explored in studies of high-$q$ quasi-free
coincidence electron scattering, specifically for polarization transfer 
observables \cite{Bof90}. It is our intent to include such effects in
treating inclusive polarization responses and to extend these studies
to exclusive (polarized target) reactions in future work. 

Drawing on our recent work \cite{Ama96} we use a new expansion in powers of 
$p/M$ (up to first-order) for the on-shell nuclear electromagnetic current,
that maintains all orders in $q/M$ and $\omega/M$ and, in addition, since we 
use relativistic kinematics in computing the momentum of the ejected nucleon,
our approach incorporates several specific classes of relativistic 
contributions. These developments permit us to compute the responses at 
high momentum transfer and accordingly results are presented for the 
CSM and PWIA (on-shell) at $q=300$--1000 MeV/c.  Comparisons have 
also been made for results in PWIA obtained with on- and off-shell 
single-nucleon currents.

In the present work our formalism has been applied to several selected 
nuclei containing a proton hole in a closed shell and special emphasis 
has been placed on $^{39}$K, being a good candidate for polarization 
measurements. Naturally it is straightforward to apply the same ideas to 
nuclei having a single proton above a closed shell and/or to odd-neutron 
cases; for brevity we have focused on proton hole examples.
The study carried out for the selected nuclei shows that:

\begin{enumerate}
\item The PWIA responses are consistently similar to the CSM with the 
exception of a more or less constant shift to higher $\omega$, of the order of 
the nucleon binding energy, which arises since in PWIA the interaction 
energy of the ejected particle is neglected.

\item The results obtained in PWIA with on-shell and off-shell (CC1) 
single-nucleon currents do not differ significantly, usually falling at 
the few percent level except in a few cases where 10\% is reached. This sets 
a scale of ``theoretical uncertainty'' for the other effects studies --- 
clearly differences found to be at the few percent level would presently 
be hard to interpret.

\item The first-order (convection and spin-orbit) terms 
of the electromagnetic current play an important role in the 
$TL$ response, while they fall typically at the few percent level 
for the other responses (excepting the $TT$ response, see below). Given 
the inevitable uncertainties arising from on-/off-shell ambiguities, these
few percent first-order effects will be hard to isolate. The $TL$ response,
however, has large enough first-order effects to make it a good 
candidate for such studies; in addition this response also shows 
measurable distortion effects. 

\item  Second-order terms in the responses are very small, 
suggesting that the first-order truncation in our formalism 
is more than sufficient for the typical momenta involved. 

\item The unprimed responses $L$, $T$, $TL$ and $TT$ are not very sensitive 
to the value of the orbital angular momentum $l$ of the hole shell, and they 
only depend of the value of $j$. In contrast, the primed responses 
$T'$ and $TL'$ present different structures for different $l$'s and equal $j$.

\item The $TT$ response is compatible with zero in PWIA, because
it is of second-order in $p/M$; however, 
it is considerably enhanced by distortion effects, since the
zeroth-order terms are nonzero in the CSM. Thus, while small, this response is
quite sensitive to details in the nuclear model of the reaction. 

\end{enumerate}

In summary, in contrast to the traditional unpolarized quasielastic responses
where distortion and first-order current terms are not very
important (except that the former produces an overall shift in energy), we 
have shown that these are in some cases essential for a proper description 
of the polarized inclusive nuclear responses. Of special interest for nuclear 
structure studies is the dramatic behaviour seen for the odd-rank $T'$ 
and $TL'$ polarized responses 
when different single-particle orbits are involved. Clearly when the 
nuclear ground state is not a one-hole configuration, but has other 
multi-particle-hole states mixed in, shells other than that at the Fermi 
surface may play a role. The fact that the polarized responses display very 
large sensitivity, even changing sign in some cases such as in going from 
$1p_{3/2}$ to $1p_{1/2}$, suggests that inclusive polarized quasielastic 
electron scattering at relatively high momentum transfers may provide an 
excellent tool for studies of near-valence nucleon momentum distributions.

\appendix

\newpage
{\LARGE {\bf Appendices}}
\section{Reduced responses for a hole nucleus}

In this appendix we show how to perform the sums over 
the angular momenta of the residual nucleus, $J_B$,
and final state, $j_f$, in the reduced responses for a nucleus whose 
ground state can be described by as an extreme one-hole configuration.
In the calculation of particle-hole excitations we must consider two kinds 
of contributions, those with two holes in different shells of 
the core, $h\ne i$, and those with $h=i$; the second is the 
relevant one for the polarization observables, as we shall see below. In 
general the final state may be written
\begin{equation}
|f\rangle = \left\{ (1-\delta_{hi})[a^{\dagger}_p
            [b^{\dagger}_hb^{\dagger}_i]_{J_B}]_{j_f}
            + \delta_{hi} \frac{1}{\sqrt{2}} [a^{\dagger}_p
            [b^{\dagger}_ib^{\dagger}_i]_{J_B}]_{j_f}
            \right\}|C\rangle,
\end{equation}
where in the $h=i$ term one has $J_B=\mbox{even}$, since there are two 
fermions (holes) in the same shell. The reduced matrix element of a 
multipole operator $\hat{T}_J$ is then given by
\begin{eqnarray}
\langle f\|\hat{T}_J\| A\rangle &=&
[J_B][j_f](-1)^{J+j_p+j_h}
\seisj{j_i}{j_f}{J}{j_p}{j_h}{J_B} \nonumber\\
&& \times \left\{ (1-\delta_{hi}) \langle p \| T_J \| h \rangle
+\sqrt{2}\delta_{hi} \langle p \| T_J \| i \rangle \right\}.
\end{eqnarray}
We write the reduced response functions ${W}^K_{\Jb}$, $K=L,T,\ldots,$ 
in the shell model as sums over responses for each hole $h$:
\begin{equation}
{W}^K_{\Jb}=\sum_h{W}^K_{\Jb h},
\end{equation}
and since the following procedures are analogous for each one of the six
responses, we illustrate the steps only for the longitudinal case.

\subsection{Reduced responses for $h\ne i$}

In the case $h\ne i$ we write for eq.~(\ref{ea1}) 
\begin{equation}
{W}^L_{\Jb h}
 =  \sum_{p}\delta(\epsilon_p-\epsilon_h-\omega)
      \sum_{JJ'}[J][J'][\Jb]\xi^+_{J'J}
      \tresj{J'}{J}{\Jb}{0}{0}{0} u^L_{J'Jph}, \label{hnei}
\end{equation}
where we have defined 
\begin{eqnarray}
    u^L_{J'Jph}
&\equiv& \sum_{j_fJ_B}(-1)^{j_i+j_f}
    \seisj{J'}{J}{\cal J}{j_i}{j_i}{j_f}
    \langle p,(hi)J_B;j_f\| \hat{M}_{J'}\| A\rangle
    \langle p,(hi)J_B;j_f\| \hat{M}_{J}\| A\rangle \nonumber\\
&=& \sum_{j_f}(-1)^{j_i+j_f}[j_f]^2
    \seisj{J'}{J}{\cal J}{j_i}{j_i}{j_f}t_{CJ'}t_{CJ}\nonumber\\
& & \times \sum_{J_B} [J_B]^2 
    \seisj{j_i}{j_f}{J'}{j_p}{j_h}{J_B}
    \seisj{j_i}{j_f}{J}{j_p}{j_h}{J_B}.
\end{eqnarray}
Here $t_{CJ}= \langle p \| M_{J} \| h\rangle$ is the Coulomb
multipole for the single-particle-hole excitation. 
Using the orthogonality of the 6-j symbols the last summation yields 
$\delta_{JJ'}\frac{1}{[J]^2}$ and hence
\begin{equation}
    u^L_{J'Jph}= \delta_{JJ'}\sum_{j_f}(-1)^{j_i+j_f}
    \seisj{J'}{J}{\cal J}{j_i}{j_i}{j_f}
    \frac{[j_f]^2}{[J]^2} t_{CJ}^2.
\end{equation}
We can also perform the sum over $j_f$ using
\begin{equation}
 \sum_{j_f}(-1)^{j_i+j_f}
 \seisj{J'}{J}{\cal J}{j_i}{j_i}{j_f} [j_f]^2
=\delta_{\Jb 0}(-1)^J[J][j_i]
\end{equation}
and therefore obtain
\begin{equation}
    u^L_{J'Jph}= \delta_{JJ'}\delta_{\Jb 0}
    (-1)^J\frac{[j_i]}{[J]} t_{CJ}^2.
\end{equation}
Evaluating the 3-j symbol in eq.~(\ref{hnei}) for ${\cal J}=0$ we obtain the 
required result for the $h\ne i$ case:
\begin{equation}
{W}^L_{\Jb h}
 = \delta_{\Jb 0}[j_i] \sum_{pJ}\delta(\epsilon_p-\epsilon_h-\omega)
   t_{CJ}^2 .
\end{equation}      
The same procedure may be followed for the other responses: in each instance
one first sums over $J_B$, getting $J=J'$, and then sums over $j_f$ obtaining
a factor $\delta_{\Jb 0}$. Since ${\cal J}=0$ corresponds to the unpolarized
result ({\it i.e.,} only $f_0^i$ occurs in eqs.~(\ref{ea5}--\ref{ea6})), no 
polarization dependence arises from the closed shell, as expected.

\subsection{Reduced responses for $h=i$}

In this case the procedure is similar, although now we do not
obtain $J=J'$ because the only allowed values for $J_B$ are 
even. In particular, for the longitudinal response we have
\begin{eqnarray}
    u^L_{J'Jpi}
&=& \sum_{j_f}(-1)^{j_i+j_f}
    \seisj{J'}{J}{\cal J}{j_i}{j_i}{j_f}t_{CJ'}t_{CJ} \nonumber\\
&&  \times \sum_{J_B=\mbox{\small even}} 2 [J_B]^2[j_f]^2 
    \seisj{j_i}{j_f}{J'}{j_p}{j_i}{J_B}
    \seisj{j_i}{j_f}{J}{j_p}{j_i}{J_B},
\end{eqnarray}
where now $t_{CJ}= \langle p \| M_{J} \| i\rangle$. 
The sum over $J_B$ in this case (see eqs.~(6.2.9) and (6.2.11) in 
ref.~\cite{Edmonds}) gives $\frac{\delta_{JJ'}}{[J]^2}-(-1)^{2j_i+J'+J}
 \seisj{j_i}{j_f}{J}{j_i}{j_p}{J'}$.
The $\delta_{JJ'}$ term just gives the same result as above and corresponds 
to the unpolarized reduced response of the complete shell $i$; the term with 
the 6-j is the only one that contributes to the polarization observables. We 
find
\begin{eqnarray}
u^L_{J'Jpi}
&=& \delta_{JJ'}\delta_{\Jb 0}
    (-1)^J\frac{[j_i]}{[J]} t_{CJ}^2    \nonumber\\
& &\kern -1cm \mbox{}+\sum_{j_f}(-1)^{j_i+j_f}
    \seisj{J'}{J}{\Jb}{j_i}{j_i}{j_f}(-1)^{J'+J}[j_f]^2
    \seisj{j_i}{j_f}{J}{j_i}{j_p}{J'}
    t_{CJ'}t_{CJ} .  \nonumber\\
\end{eqnarray}
For the second term here we use eq.~(6.2.11) in ref.~\cite{Edmonds}, 
\begin{equation}
\sum_{j_f}(-1)^{j_i+j_f}[j_f]^2
\seisj{J'}{J}{\Jb}{j_i}{j_i}{j_f}
\seisj{j_i}{j_f}{J}{j_i}{j_p}{J'}
=(-1)^{j_i-\Jb-j_p}\seisj{J'}{J}{\Jb}{j_i}{j_i}{j_p},
\end{equation}
yielding
\begin{eqnarray}
u^L_{J'Jpi}
&=& \delta_{JJ'}\delta_{\Jb 0}
    (-1)^J\frac{[j_i]}{[J]} t_{CJ}^2  \nonumber\\
& & \kern -1cm \mbox{}-(-1)^{j_i+j_p}(-1)^{J+J'+\Jb}
    \seisj{J'}{J}{\Jb}{j_i}{j_i}{j_p}
    t_{CJ'}t_{CJ}.
\end{eqnarray}
Finally, the total longitudinal reduced response can be cast in the form
\begin{eqnarray}
{W}^L_{\Jb i}
& = & \delta_{\Jb 0}[j_i] \sum_{pJ}\delta(\epsilon_p-\epsilon_i-\omega)
      t_{CJ}^2 \nonumber\\
&   & \kern -1.5cm \mbox{} 
    - \sum_{pJJ'}\delta(\epsilon_p-\epsilon_i-\omega)
      (-1)^{j_i+j_p+\Jb}[J][J'][\Jb] \nonumber\\
&&   \times   \tresj{J'}{J}{\Jb}{0}{0}{0}
      \seisj{J'}{J}{\Jb}{j_i}{j_i}{j_p}
      \xi^+_{J'J}t_{CJ'}t_{CJ}, 
\end{eqnarray}      
with similar results for $T$, $TL$, {\it etc.}

\section{Multipoles of the spin-orbit charge density}

Here we present the formalism used in the computation of the
Coulomb reduced matrix elements of  the spin-orbit term.
We begin with the matrix element of the spin-orbit charge
density operator (see ref.~\cite{Ama96}) between nucleon plane-wave states,
\begin{equation}
  \langle \np's'|\rho_{so}(\nq,\omega)|\np,s\rangle
  = \delta(\np+\nq-\np')
    \frac{2G_M-G_E}{\sqrt{1+\tau}}
    \frac{i}{4M^2}(\nq\times\np)\cdot\nsigma_{s's}.
\end{equation}
It is convenient to define a ``bare'' operator $\overline{\rho}_{so}$ extracting
the form factor $2G_M-G_E$ and the kinematical factor 
$\sqrt{1+\tau}$:
\begin{eqnarray}
    \rho_{so}(\nq) 
&\equiv& \frac{2G_M-G_E}{\sqrt{1+\tau}}
    \overline{\rho}_{so}(\nq) \\
    \langle \np's'|\overline{\rho}_{so}(\nq)|\np,s\rangle
&=& \delta(\np+\nq-\np')
    \frac{i}{4M^2}(\nq\times\np)\cdot\nsigma_{s's},
\end{eqnarray}
where now $\overline{\rho_{so}}$ is independent of the energy transfer
$\omega$. The coordinate-space expression for  
this operator is
\begin{equation}
\overline{\rho}_{so}(\nr)
                  = \frac{i}{8M^2}\nabla\times
                    \left[\delta(\nr-\nr_1)\nabla_1
                          +\nabla_1\delta(\nr-\nr_1)
                    \right]\cdot\nsigma,
\end{equation}
where $\nr_1$ denotes the position operator of the nucleon
and $-i\nabla_1=\np_1$ its momentum operator. The symmetrized form 
$i(\delta\nabla+\nabla\delta)$ ensures the hermiticity of the operator.

The Coulomb multipole operator for the spin-orbit charge density is written
\begin{equation}
M_{JM}(q)=\frac{2G_M-G_E}{\sqrt{1+\tau}}\overline{M}_{JM}(q),
\end{equation}
where the ``bare'' multipole operator $\overline{M}_{JM}(q)$
is defined from the ``bare'' charge-density $\overline{\rho}_{so}$
in the usual way
\begin{equation}
\overline{M}_{JM}(q) = \int d^3r\, j_J(qr)Y_{JM}(\hat{\nr})
                      \overline{\rho}_{so}(\nr).
\end{equation}
Inserting the coordinate-space expression for $\overline{\rho}_{so}$
we obtain 
\begin{equation}
\overline{M}_{JM}(q)= -\frac{i}{4M^2}
                     [\nabla_1j_J(qr_1)Y_{JM}(\hat{\nr_1})]
                     \times\nabla_1\cdot\nsigma,
\end{equation}
where the first gradient $\nabla_1$ only operates on the
functions within the brackets. For simplicity, in the following
we shall call $\nr$ the coordinate of the nucleon (instead
of $\nr_1$). The gradient of a spherical Bessel function times a spherical
harmonic may be expressed in terms of vector spherical harmonics (see 
eq.~(5.9.17) in ref.~\cite{Edmonds}), yielding
\begin{equation} \label{coulomb-vector}
\overline{M}_{JM}(q)
  = -\frac{i}{4M^2}
     \frac{q}{[J]}\sum_{s=\pm1}
     \sqrt{J+\delta_{s1}}\,\,
     j_{J'}(qr)[\nY_{J'JM}(\hat{\nr})\times\nabla]\cdot\nsigma,
\end{equation}
where $J'=J+s$ and as usual $[J]=\sqrt{2J+1}$.

In order to compute the reduced matrix elements of the Coulomb multipole
operator $\overline{M}_{JM}$, it is convenient to write it as a linear 
combination of the basic irreducible tensor operators 
\begin{equation}
{\cal D}_{J'LJM}\equiv
 \left[\left[Y_{J'}(\hat{\nr})\otimes\nabla\right]_L
       \otimes\sigma
 \right]_{JM}.
\end{equation}
To this end we use the facts that for any vector operator $\nA$ one has
$\nY_{lJM}\cdot\nA=\left[Y_l\otimes A\right]_{JM}$ and for any 
two vector operators $\nA$ and $\nB$ one has
$(\nA\times\nB)_{\alpha}=-i\sqrt{2}[A\otimes B]_{1\alpha}$. One then has
\begin{equation}
[\nY_{J'JM}\times\nabla]\cdot\nsigma
=\nY_{J'JM}\cdot[\nabla\times\nsigma]
=-i\sqrt{2}\left[ Y_{J'}\otimes
                  \left[\nabla\otimes\sigma\right]_1
           \right]_{JM},
\end{equation}
or, using a 6-j symbol to recouple the angular momenta, one has
\begin{equation}
[\nY_{J'JM}\times\nabla]\cdot\nsigma=
i\sqrt{6}\sum_L[L]\seisj{J'}{1}{L}{1}{J}{1}
\left[\left[Y_{J'}\otimes\nabla\right]_L\otimes\sigma\right]_{JM}.
\end{equation}
Inserting this expression in eq.~(\ref{coulomb-vector}) we obtain 
\begin{equation}
\overline{M}_{JM}(q)=\frac{\sqrt{6}}{4M^2}\frac{q}{[J]}
\sum_{s=\pm 1}\sqrt{J+\delta_{s1}} S_{J'JM}(q),
\end{equation}
where the operator $S_{J'JM}(q)$ is defined as
\begin{equation}
S_{J'JM}(q)\equiv \sum_{L=J,J'}[L]\seisj{J'}{1}{L}{1}{J}{1}
j_{J'}(qr){\cal D}_{J'LJM}.
\end{equation}
Note that the sum over $L$ has only two nonzero terms, $L=J,J'$, because the
two triangle relations $J-1 \leq L \leq J+1$ and $J'-1\leq L \leq J'+1$
must hold simultaneously.

In evaluating the matrix elements of $S_{J'JM}(q)$ it is useful to treat 
spin, orbital angular momentum and radial dependences separately, since 
our single-particle wave functions are labeled 
$|p\rangle=|n_p\frac12 l_p;j_pm_p\rangle$. We must compute the reduced 
matrix element taken between particle and hole states:
\begin{equation}\label{27}
\langle p\|S_{J'JM}(q)\| h\rangle=
\sum_{L=J,J'}[L]\seisj{J'}{1}{L}{1}{J}{1}
\langle p\|j_{J'}(qr){\cal D}_{J'LJ}\| h\rangle.
\end{equation}
The spin reduced matrix element is given by 
$\langle\frac12\|\sigma\|\frac12\rangle =\sqrt{6}$ and for the angular 
reduced matrix element we have
\begin{eqnarray}
         \langle l_p\|\left[Y_{J'}\otimes\nabla\right]_L\| l_h\rangle
&=&      \frac{(-)^{l_h+L}}{\sqrt{4\pi}}[L][l_p][J']
         \sum_{s_h=\pm 1} s_h\sqrt{l_h+\delta_{s_h,1}}[L_h]
         \seisj{J'}{1}{L}{l_h}{l_p}{L_h}\nonumber\\
&\times& \tresj{L_h}{J'}{l_p}{0}{0}{0}
         \left(\frac{d}{dr}-s_h\frac{l_h+\delta_{s_h,-1}}{r}\right),
\end{eqnarray}
defining $L_h=l_h+s_h$ and using the fact that
\begin{equation}
    \langle l\|\nabla\| l_h\rangle
= \sum_{s_h=\pm 1} s_h\sqrt{l_h+\delta_{s_h,1}}\delta_{lL_h}
    \left(\frac{d}{dr}-s_h\frac{l_h+\delta_{s_h,-1}}{r}\right)
     \label{32}
\end{equation}
together with eqs.~(5.4.5) and (7.1.1) in ref.~\cite{Edmonds}.
Using this result we arrive at the following expression for the required
matrix elements:
\begin{eqnarray}
\langle p\|j_{J'}{\cal D}_{J'LJ}\| h\rangle
&=& (-)^{l_h+J+1}[j_p][j_h][J][J'][L][l_p]\frac{\sqrt{6}}{\sqrt{4\pi}}
    \nuevej{\frac12}{l_p}{j_p}{\frac12}{l_h}{j_h}{1}{L}{J}\nonumber\\ 
&&   \times \sum_{s_h=\pm1}s_h\sqrt{l_h+\delta_{s_h,-1}}[L_h]
       \seisj{J'}{1}{L}{l_h}{l_p}{L_h}
         \tresj{L_h}{J'}{l_p}{0}{0}{0}\nonumber\\
&&     \times    \int_0^{\infty}dr\,r^2 R_p\, j_{J'}(qr)
         \left(\frac{d}{dr}-s_h\frac{l_h+\delta_{s_h,-1}}{r}\right)R_h,
\end{eqnarray}
where $R_p(r)$ and $R_h(r)$ are the radial wave functions. Making explicit 
the sum over $L$, the matrix element in eq.~(\ref{27}) then becomes
\begin{eqnarray}
\langle p\|S_{J'JM}(q)\| h\rangle 
&  =   & \nonumber\\
&& \kern -4cm [J]\seisj{J'}{1}{J}{1}{J}{1}(-)^{l_h+J+1}
         [j_p][j_h][J][J'][J][l_p]
         \frac{\sqrt{6}}{\sqrt{4\pi}}
         \nuevej{\frac12}{l_p}{j_p}{\frac12}{l_h}{j_h}{1}{J}{J}
         \nonumber\\
&& \kern -4cm \times\sum_{s_h=\pm1}s_h\sqrt{l_h+\delta_{s_h,-1}}[L_h]
         \seisj{J'}{1}{J}{l_h}{l_p}{L_h}
         \tresj{L_h}{J'}{l_p}{0}{0}{0} 
         \nonumber\\
&& \kern -4cm \times\int_0^{\infty}dr\,r^2 R_p\, j_{J'}(qr)
         \left(\frac{d}{dr}-s_h\frac{l_h+\delta_{s_h,-1}}{r}\right)R_h
         \nonumber\\
& &\kern -4cm + [J']\seisj{J'}{1}{J'}{1}{J}{1}
         (-)^{l_h+J+1}[j_p][j_h][J][J'][J][l_p]
         \frac{\sqrt{6}}{\sqrt{4\pi}}
         \nuevej{\frac12}{l_p}{j_p}{\frac12}{l_h}{j_h}{1}{J'}{J} 
         \nonumber\\
&& \kern -4cm \times\sum_{s_h=\pm1}s_h\sqrt{l_h+\delta_{s_h,-1}}[L_h]
         \seisj{J'}{1}{J'}{l_h}{l_p}{L_h}
         \tresj{L_h}{J'}{l_p}{0}{0}{0} 
         \nonumber\\
&& \kern -4cm \times\int_0^{\infty}dr\,r^2 R_p\, j_{J'}(qr)
         \left(\frac{d}{dr}-s_h\frac{l_h+\delta_{s_h,-1}}{r}\right)R_h.
\end{eqnarray}
Moreover, the product of a  6-$j$ with the 3-$j$ can be evaluated using
\begin{eqnarray}
      \seisj{J'}{1}{J}{l_h}{l_p}{L_h}
      \tresj{L_h}{J'}{l_p}{0}{0}{0} 
& = & \nonumber\\
&&\kern -5cm \frac{P^+_{l_p+l_h+J}}{[L_h][l_h][J][J']}
      \left\{ [(l_h+\delta_{s_h,-1})(J+\delta_{s,-1})]^{1/2}
              \tresj{l_h}{J}{l_p}{1}{-1}{0} 
      \right. \nonumber\\
&&\kern -5cm \left.\mbox{}-s_hs[(l_h+\delta_{s_h,1})(J+\delta_{s,1})]^{1/2}
              \tresj{l_h}{J}{l_p}{0}{0}{0}  
     \right\}
\end{eqnarray}
\begin{equation}
     \seisj{J'}{1}{J'}{l_h}{l_p}{L_h}
     \tresj{L_h}{J'}{l_p}{0}{0}{0}  = 
   \frac{P^+_{l_p+l_h+J}}{[L_h][l_h][J']}(l_h+\delta_{s_h,-1})^{1/2}
     \tresj{l_h}{J'}{l_p}{1}{-1}{0} .
\end{equation}
We may then perform the sums over $s_h$ and, after some algebra, write 
the matrix element in the following way:
\begin{eqnarray}
\langle p \| S_{J'J}(q) \| h\rangle
&=& A_{J'J}(ph)
  \int_0^{\infty}dr\, r R_{p}(r)j_{J'}(qr)R_{h}(r)\nonumber\\
&&+ B_{J'J}(ph)
  \int_0^{\infty}dr\, r^2 R_{p}(r)j_{J'}(qr)\frac{dR_{h}}{dr},
\end{eqnarray}
where we have defined the coupling factors:
\begin{eqnarray}
A_{J'J}(ph) &=& P^+_{l_p+l_h+J}a_{phJ}[J]
                \sqrt{(J+\delta_{s,-1})l_h(l_h+1)}\nonumber\\
&&       \times         \seisj{J'}{1}{J}{1}{J}{1}
                \nuevej{\frac12}{l_p}{j_p}{\frac12}{l_h}{j_h}{1}{J}{J}
                \tresj{l_h}{J}{l_p}{1}{-1}{0}\nonumber\\
            & &+P^+_{l_p+l_h+J}a_{phJ}[J']^2
                \sqrt{l_h(l_h+1)} \nonumber\\
&&        \times        \seisj{J'}{1}{J'}{1}{J}{1}
                \nuevej{\frac12}{l_p}{j_p}{\frac12}{l_h}{j_h}{1}{J'}{J}
                \tresj{l_h}{J'}{l_p}{1}{-1}{0}\\
B_{J'J}(ph) &=& P^+_{l_p+l_h+J}a_{phJ}[J]
                s\sqrt{J+\delta_{s,1}} \nonumber\\
&&        \times        \seisj{J'}{1}{J}{1}{J}{1}
                \nuevej{\frac12}{l_p}{j_p}{\frac12}{l_h}{j_h}{1}{J}{J}
                \tresj{l_h}{J}{l_p}{0}{0}{0}
\end{eqnarray}
with
\begin{equation}
a_{phJ}\equiv\sqrt{\frac{3}{2\pi}}
                (-1)^{l_p}[l_p][l_h][j_p][j_h][J].
\end{equation}
These lead to the required matrix elements of the Coulomb 
operator $\overline{M}_J$:
\begin{equation}
\langle p \| \overline{M}_{J}(q) \| h\rangle
= \frac{\sqrt{6}}{4M^2}\frac{q}{[J]}
  \left[ \sqrt{J}\langle p \| S_{J-1,J}(q) \| h\rangle
        +\sqrt{J+1}\langle p \| S_{J+1,J}(q) \| h\rangle
  \right].
\end{equation}

Finally, we note the following useful identities to assist in computing 
quasielastic responses where the sums over multipoles involve large numbers 
of coupling coefficients.  To speed up the calculation it is convenient to 
reduce the order of the $n$-$j$ coefficients whenever possible employing 
the following:
\begin{eqnarray}
\nuevej{\frac12}{l_p}{j_p}{\frac12}{l_h}{j_h}{1}{J}{J}
\tresj{l_h}{J}{l_p}{0}{0}{0}
& = & \nonumber\\
&& \kern -3cm P^+_{l_p+l_h+J}\frac{(-1)^{j_p+l_p-1/2}}{\sqrt{6}[l_p][l_h][J]}
      \frac{\kappa_p-\kappa_h}{\sqrt{J(J+1)}}
      \tresj{j_p}{j_h}{J}{-\frac12}{\frac12}{0}       \\
\nuevej{\frac12}{l_p}{j_p}{\frac12}{l_h}{j_h}{1}{J}{J}
\tresj{l_h}{J}{l_p}{1}{-1}{0}
& = & \frac{1}{2\sqrt{6}[l_p][l_h][J]}
      \left\{
      \left[\frac{\kappa_h+1}{\kappa_h}\right]^{1/2}
      \tresj{j_h}{j_p}{J}{-\frac12}{\frac12}{0} 
      \right. \nonumber\\
& & \kern -1cm + \frac{(-1)^{j_p+l_p+1/2}}{\sqrt{J(J+1)}}
      \left[\frac{\kappa_h+1}{\kappa_h}\right]^{1/2}
      \tresj{j_h}{j_p}{J}{-\frac12}{-\frac12}{1} \nonumber\\
& & \kern -1cm + \frac{(-1)^{j_h+l_h+1/2}}{\sqrt{J(J+1)}}
      \left[\frac{\kappa_h-1}{\kappa_h}\right]^{1/2}
      \tresj{j_h}{j_p}{J}{-\frac32}{\frac12}{1} \nonumber\\
& & \kern -4cm \left.
    + (-1)^{j_p+j_h+J}
      \left[\frac{(J-1)(J+2)}{J(J+1)}
            \frac{\kappa_h-1}{\kappa_h}\right]^{1/2}
      \tresj{j_h}{j_p}{J}{-\frac32}{-\frac12}{2} 
    \right\}    \\
\nuevej{\frac12}{l_p}{j_p}{\frac12}{l_h}{j_h}{1}{J'}{J}
\tresj{l_h}{J'}{l_p}{1}{-1}{0} 
& = & \nonumber\\
&&\kern -5cm
      \frac{1}{2\sqrt{3}[l_p][l_h][J][J']}
      \left\{
      \left[\frac{(J-s)(J-s+1)}{2J+s+1}
            \frac{\kappa_h+1}{\kappa_h}\right]^{1/2}
      \tresj{j_h}{j_p}{J}{\frac32}{\frac12}{-2} 
      \right. \nonumber\\
&&\kern -5cm
  + (-1)^{l_p+j_p-1/2}\frac{s}{2}
      \left[\frac{(2J+3+s)(2J+s-1)}{2J+s+1}
            \frac{\kappa_h-1}{\kappa_h}\right]^{1/2}
      \tresj{j_h}{j_p}{J}{\frac32}{-\frac12}{-1} 
      \nonumber\\
&&\kern -5cm
  + (-1)^{l_h+j_h-1/2}\frac{s}{2}
      \left[\frac{(2J+3+s)(2J+s-1)}{2J+s+1}
            \frac{\kappa_h+1}{\kappa_h}\right]^{1/2}
      \tresj{j_h}{j_p}{J}{\frac12}{\frac12}{-1} 
       \nonumber\\
&&\kern -5cm
  + \left.(-1)^{j_p+j_h+J+1}
      \left[\frac{(J+s)(J+s+1)}{2J+s+1}
            \frac{\kappa_h+1}{\kappa_h}\right]^{1/2}
      \tresj{j_h}{j_p}{J}{\frac12}{-\frac12}{0} 
      \right\},
\end{eqnarray}
where as before $\kappa_i= (-1)^{j_i+l_i+1/2}(j_i+1/2)$.

\section{Polarized responses for a hole nucleus in PWIA}

In this appendix we give some of the details of the calculation of the
polarized momentum distribution and inclusive responses 
for a hole nucleus in PWIA omitted from sect.~\ref{PolPWIA}. 
In order to compute the momentum distribution of the shell $h$,
\begin{equation}
n(\np)_{r'r}=
\sum_{m_hm'_i}
\langle B|a_{\np r}|A\rangle^*
\langle B|a_{\np r'}|A\rangle,
\end{equation}
we need to compute the matrix element of the annihilation
operator $a_{\np r}$ between the initial and the residual 
nuclear states (see eqs.~(\ref{eqn58},\ref{eqn59})):
\begin{equation}
\langle B | a_{\np r}|A\rangle
= \sum_{m_i}
  \langle C|b_{im'_i}b_{hm_h}a_{\np r} b^{\dagger}_{im_i}|C\rangle
  {\cal D}_{m_i i}^{(i)}(\Omega^*).
\end{equation}
To accomplish this, we expand the destruction operator in the plane-wave basis 
in terms of the shell model basis,
\begin{equation}
a_{\np r}=
\sum_{l<F}\langle \np, r|\tilde{l}\rangle b^{\dagger}_l
+\sum_{\alpha>F}\langle \np, r|\alpha\rangle a_{\alpha},
\end{equation}
where $|\tilde{l}\rangle=|\widetilde{j_lm_l}\rangle=
(-1)^{j_l+m_l}|j_l-m_l\rangle$ is the time inversion of the 
single-particle state $|l\rangle$. Using the commutation relations
between particle and hole operators, we obtain for the matrix element
\begin{equation}
\langle B | a_{\np r}|A\rangle
={\cal D}_{m'_ii}^{(i)}\langle\np r|\widetilde{hm_h}\rangle
-\delta_{hi}{\cal D}_{m_hi}^{(i)}\langle\np r|\widetilde{im'_i}\rangle.
\end{equation}
In the case $h\ne i$ we have
\begin{equation} 
\langle B| a_{\np r}|A\rangle = {\cal D}_{m'_ii}^{(i)}
                                \langle\np r| \widetilde{h m_h}\rangle,
\end{equation}
and thus we sum over third
components $m_h$ and use $\sum_{m'_i}|{\cal D}_{m'_ii}^{(i)}|^2=1$
to obtain the momentum distribution of the complete shell $h$:
\begin{equation}
n(\np)_{r'r}=n^{(h)}_{r'r}(\np)_{\rm unpol}=
\sum_{m_h}\langle\np r'|hm_h\rangle\langle hm_h|\np r'\rangle.
\end{equation}
In the case $h=i$, $m_h=m''_i$ and we have
\begin{equation}
\langle B|a_{\np r}|A\rangle=
{\cal D}_{m'_ii}^{(i)}\langle\np r|\widetilde{im''_i}\rangle
-{\cal D}_{m''_ii}^{(i)}\langle\np r|\widetilde{im'_i}\rangle
\end{equation}
An important aspect here is that in the sum over 
$m_i'$, $m''_i$ we must divide by a factor two
to avoid double-counting in the antisymmetrized states, 
$|B\rangle=|i^{-1}m''_i,i^{-1}m'_i\rangle$.
Taking this fact into account we obtain for the momentum distribution
\begin{eqnarray}
n(\np)_{r'r}&=&\sum_{m_i}\langle\np r'|im_i\rangle
                         \langle im_i|\np r\rangle
              -\sum_{m'_i m''_i}{\cal D}_{m'_ii}^{(i)*}
                       {\cal D}_{m''_ii}^{(i)}
                       \langle\np r'|\widetilde{im'_i}\rangle
                       \langle \widetilde{im''_i}|\np r\rangle\nonumber\\
&=& n^{(i)}_{r'r}(\np)_{\rm unpol}
   -n^{(i)}_{r'r}(\np,\Omega^*)_{\rm hole},
\end{eqnarray}
that is, the (unpolarized) 
momentum distribution of the complete shell $i$ minus
the (polarized) momentum distribution of the hole,
\begin{equation}
n^{(i)}_{r'r}(\np,\Omega^*)_{\rm hole}
=  \sum_{m'_i m''_i}{\cal D}_{m'_ii}^{(i)*}
                       {\cal D}_{m''_ii}^{(i)}
                       \langle\np r'|\widetilde{im'_i}\rangle
                       \langle \widetilde{im''_i}|\np r\rangle.
\end{equation}
For the computation of the unpolarized and polarized momentum
distribution we must sustitute the matrix element
\begin{equation}
\langle\np r|jm\rangle = i^{-l}\sum_M Y_{lM}(\hat{\np})
                         \langle{\textstyle\frac12} r lM| jm\rangle
                         \tilde{R}(p)
\end{equation}
and perform the sums over third components using Racah algebra. 

These basic PWIA developments can be found for example in ref.~\cite{Cab94} 
where a general expression is given and thus we do not repeat the details
here, but only quote the result for the particular model of present interest.
The polarized momentum distribution of a hole with quantum numbers
$l,j$ can be written 
\begin{eqnarray}
    n^{(j)}_{rr'}(\np,\Omega^*)_{\rm hole}
&=& [l]^2[j]^2    \tilde{R}(p)^2
    \sum_{SM}[S](-1)^{r'-1/2} 
    \tresj{\frac12}{\frac12}{S}{r}{-r'}{M}\nonumber\\
& & \kern -3cm \times\sum_{\Jb\Jb'}(-1)^{l+\Jb}f_{\Jb}^{(j)}[\Jb']
    \nuevej{\frac12}{\frac12}{S}{l}{l}{\Jb'}{j}{j}{\Jb}
    \tresj{\Jb'}{l}{l}{0}{0}{0}
    [Y_{\Jb}(\Omega^*)Y_{\Jb'}(\hat{\np})]_{SM},
\end{eqnarray}
where the coupling of two spherical harmonics with different
angles is defined by
\begin{equation}
[Y_{\Jb}(\Omega^*)Y_{\Jb'}(\hat{\np})]_{SM}
\equiv \sum_{\Mb\Mb'}\langle \Jb\Mb\Jb'\Mb'|SM\rangle
  Y_{\Jb\Mb}(\Omega^*)Y_{\Jb'\Mb'}(\hat{\np}).
\end{equation}
First, note that in this equation we have $S=0,1$, and hence may write 
$n_{rr'}$ as the sum of scalar ($S=0$) plus vector ($S=1$)
terms, $n=n^S+n^V$. Obviously $n^S$ and $n^V$ are related to the
scalar and vector momentun distributions introduced in sect.~\ref{PolPWIA}
by
\begin{eqnarray}
n^S_{rr'}&=&\frac12\delta_{rr'}M^S\\
n^V_{rr'}&=&\frac12\langle r |\nM^V\cdot\nsigma|r'\rangle.
\end{eqnarray}
Second we note from the second 3-$j$ 
that $\Jb'=\rm even$ and as a consequence it is easily seen by
permutation of the first two columns of the 9-$j$ that the
result is zero unless $S+\Jb=\rm even$. 

For $S=0$ we have $\Jb=\Jb'=\rm even$. In this case we can 
use eqs.~(3.7.9) and (6.4.14) in ref.~\cite{Edmonds} together with the 
following relation
\begin{equation}
    \seisj{l}{l}{\Jb}{j}{j}{\frac12}
    \tresj{\Jb}{l}{l}{0}{0}{0}
= -P^+_{\Jb}\frac{1}{[l]^2}
    \tresj{j}{j}{\Jb}{\frac12}{-\frac12}{0}
\end{equation}
to obtain
\begin{equation}
n^S_{rr'}=\delta_{rr'}[j]^2\tilde{R}(p)^2
          \sum_{\Jb}P^+_{\Jb}f^{(j)}_{\Jb}
          \frac{(-1)^{j-1/2}}{2}
          \tresj{j}{j}{\Jb}{\frac12}{-\frac12}{0}
    [Y_{\Jb}(\Omega^*)Y_{\Jb}(\hat{\np})]_{0},
\end{equation}
from which we immediately arrive at eq.~(\ref{66}).
In the $S=1$ case we have $\Jb=\rm odd$ and $\Jb'=\Jb\pm 1$.
We use the following expression for the product of the 9-$j$
with the 3-$j$
\begin{eqnarray}
\nuevej{\frac12}{\frac12}{1}{l}{l}{\Jb'}{j}{j}{\Jb}
\tresj{\Jb'}{l}{l}{0}{0}{0} = \nonumber\\
&& \kern -3cm 
           P^-_{\Jb}\frac{(-1)^{j+l+1/2}}{\sqrt{6}[l]^2[\Jb][\Jb']}
           \frac{2\kappa+s\Jb+\delta_{s1}}{\sqrt{\Jb+\delta_{s1}}}
           \tresj{j}{j}{\Jb}{\frac12}{-\frac12}{0}
\end{eqnarray}
to obtain for the vector part of the momentum distribution
\begin{eqnarray}
n^V_{rr'}&=&\sqrt{2}(-1)^{r'-1/2}\sum_{\mu}
          \tresj{\frac12}{\frac12}{1}{r}{-r'}{\mu}\nonumber\\
&&       \times   \sum_{\Jb}\sum_{\Jb'=\Jb\pm 1}
          P^-_{\Jb}f_{\Jb}^j[j]^2|\tilde{R}(p)|^2 A_{\Jb\Jb'}
          [Y_{\Jb}(\Omega^*)Y_{\Jb'}(\hat{\np})]_{1\mu},
\end{eqnarray}
where we have used the definitions eqs.~(\ref{68}--\ref{70}). From this
result it is straightforward to obtain eq.~(\ref{67}) for the 
vector momentum distribution.

To obtain the inclusive responses for a one-hole nucleus in PWIA we begin 
with the general expressions in eqs.~(\ref{90a},\ref{90b}), 
using the single-nucleon responses given in eqs.~(\ref{SNR-L}-\ref{SNR-T'})
and the scalar and vector momentum distributions given in 
eqs.~(\ref{66},\ref{67}). The integral over $\phi$ in (\ref{90a},\ref{90b}) 
can be performed analytically in all cases. 
First we define the functions $y_{lm}(\theta)$ through 
$Y_{lm}(\theta,\phi)\equiv y_{lm}(\theta){\rm e}^{im\phi}$.
The integrals that we need for the computation 
of the unprimed responses are the following:
\begin{equation}
    \int_{0}^{2\pi}{\rm d}\phi\, 
    [Y_{\Jb}(\Omega^*)Y_{\Jb}(\hat{\np})]_0\cos {\cal M}\phi
=   \frac{2\pi}{[\Jb]}y_{\Jb \cal M}(\theta^*)y_{\Jb \cal M}(\theta)
    \cos{\cal M}\phi^*,
\end{equation}
with  ${\cal M}=0,1,2$. The single-nucleon  $L$ and $T$ responses
are independent of $\phi$ and so we have ${\cal M}=0$:
\begin{eqnarray}
R^L &=& \frac{4\pi M}{q}
        \sum_{\Jb}P^+_{\Jb}f^i_{\Jb}A_{\Jb}
        \frac{[j_i]^2}{[\Jb]}y_{\Jb 0}(\theta^*)
        \int_{|p'-q|}^{p'+q}{\rm d}p\, p
        (\rho_c^2+\rho_{so}^2\delta^2)
        |\tilde{R}_i(p)|^2
        y_{\Jb 0}(\theta)\nonumber\\
  & & \\
R^T &=& \frac{4\pi M}{q}
        \sum_{\Jb}P^+_{\Jb}f^i_{\Jb}A_{\Jb}
        \frac{[j_i]^2}{[\Jb]}y_{\Jb 0}(\theta^*)
        \int_{|p'-q|}^{p'+q}{\rm d}p\, p
        (2J_m^2+J_c^2\delta^2)
        |\tilde{R}_i(p)|^2
        y_{\Jb 0}(\theta). \nonumber\\
  & & 
\end{eqnarray}
Next, the single-nucleon $TL$ response is proportional to $\cos\phi$ and
so we have ${\cal M}=1$:
\begin{eqnarray}
R^{TL} &=&    4\pi\frac{M}{q}
            \sum_{\Jb}P^+_{\Jb}f^i_{\Jb}A_{\Jb}
            \frac{[j_i]^2}{[\Jb]}y_{\Jb 1}(\theta^*)\cos\phi^*\nonumber\\
&&      \times      \int_{|p'-q|}^{p'+q}{\rm d}p\, p
            2\sqrt{2}(\rho_cJ_c+\rho_{so}J_m)\delta
            |\tilde{R}_i(p)|^2
            y_{\Jb 1}(\theta).
\end{eqnarray}
Finally, the single-nucleon $TT$ response is proportional to 
$\cos2\phi$ and so we have the integral for ${\cal M}=2$:
\begin{equation}
R^{TT} =   -4\pi\frac{M}{q}
            \sum_{\Jb}P^+_{\Jb}f^i_{\Jb}A_{\Jb}
            \frac{[j_i]^2}{[\Jb]}y_{\Jb 2}(\theta^*)\cos2\phi^*
            \int_{|p'-q|}^{p'+q}{\rm d}p\, p
            J_c^2\delta^2
            |\tilde{R}_i(p)|^2
            y_{\Jb 2}(\theta).
\end{equation}

For the primed responses we first define a real vector
$\nX^{\Jb\Jb'}(\Omega^*,\hat{\np})$ by its spherical components
\begin{equation}
X^{\Jb\Jb'}_{\alpha}(\Omega^*,\hat{\np})
\equiv [Y_{\Jb}(\Omega^*)Y_{\Jb'}(\hat{\np})]_{1\alpha}
\end{equation}
and then we can write the vector momentum distribution in eq.~(\ref{67}) as
a linear combination of the vectors $\nX^{\Jb\Jb'}$:
\begin{equation}
\nM^V= \frac{2}{\sqrt{3}}\sum_{\Jb}\sum_{\Jb'=\Jb\pm1}
        P^-_{\Jb}f^i_{\Jb}A_{\Jb\Jb'}[j_i]^2
        |\tilde{R}_i(p)|^2
        \nX^{\Jb\Jb'}.
\end{equation}
It is straightforward to compute the scalar products between the 
vector momentum distribution $\nM^V$ and the vector single-nucleon 
response $\nw_V^K$. To obtain the azimuthal integrals we need the following
results for integrals of components of $\nX$:
\begin{equation}
\int_{0}^{2\pi}{\rm d}\phi\, X_1 
= -2\pi\sqrt{2}\langle\Jb1\Jb'0|11\rangle
    y_{\Jb1}(\theta^*)y_{\Jb'0}(\theta)\cos\phi^*
\end{equation}
\begin{eqnarray}
\int_{0}^{2\pi}{\rm d}\phi\,\sin\phi(\cos\phi\, X_2-\sin\phi\, X_1) 
&=& \pi\sqrt{2} y_{\Jb1}(\theta^*)\cos\phi^*\nonumber\\
&& \kern -4cm  \times[\langle\Jb-1\Jb'2|11\rangle y_{\Jb'2}(\theta)+
    \langle\Jb1\Jb'0|11\rangle y_{\Jb'0}(\theta)]
\end{eqnarray}
\begin{equation}
\int_{0}^{2\pi}{\rm d}\phi\,\cos\phi\, X_3
= -2\pi y_{\Jb1}(\theta^*)\cos\phi^*
    \langle\Jb1\Jb'-1|10\rangle y_{\Jb'1}(\theta),
\end{equation}
from which we obtain
\begin{eqnarray}
R^{TL'}&=&  4\pi\frac{M}{q}
            \sum_{\Jb}\sum_{\Jb'\pm1}P^-_{\Jb}f^i_{\Jb}A_{\Jb\Jb'}
            [j_i]^2\sqrt{\frac{2}{3}}
            y_{\Jb 1}(\theta^*)\cos\phi^*
            \int_{|p'-q|}^{p'+q}{\rm d}p\, p
            |\tilde{R}_i(p)|^2\nonumber\\
&& \times \left\{-2\sqrt{2}\langle\Jb 1 \Jb' 0 | 11\rangle
          y_{\Jb' 0}(\theta)\rho_cJ_m
   \right. \nonumber\\
&&  +\sqrt{2}[
    \langle\Jb -1 \Jb' 2 | 11\rangle y_{\Jb' 2}(\theta)
   +\langle\Jb  1 \Jb' 0 | 11\rangle y_{\Jb' 0}(\theta)
   ]\rho_{so}J_c\delta^2
    \nonumber\\
&& +2\langle\Jb 1 \Jb' -1 | 10\rangle
          y_{\Jb' 1}(\theta)\rho_{so}J_m\delta
   \Big\}. 
\end{eqnarray}
For the response $T'$ we need the following integrals:
\begin{eqnarray}
\int_{0}^{2\pi}{\rm d}\phi\,(\cos\phi\, X_1+\sin\phi\, X_2) 
&=& -2\pi\sqrt{2}\langle\Jb0\Jb'1|11\rangle
    y_{\Jb0}(\theta^*)y_{\Jb'1}(\theta)\\
\int_{0}^{2\pi}{\rm d}\phi\, X_3 
&=& 2\pi\langle\Jb0\Jb'0|10\rangle
    y_{\Jb0}(\theta^*)y_{\Jb'0}(\theta),
\end{eqnarray}
from which we have
\begin{eqnarray}
R^{T'}&=& -4\pi\frac{M}{q}
            \sum_{\Jb}\sum_{\Jb'\pm1}P^-_{\Jb}f^i_{\Jb}A_{\Jb\Jb'}
            [j_i]^2\frac{2}{\sqrt{3}}
            y_{\Jb 0}(\theta^*)
            \int_{|p'-q|}^{p'+q}{\rm d}p\, p
            |\tilde{R}_i(p)|^2\nonumber\\
&& \times \left\{\sqrt{2}\langle\Jb 0 \Jb' 1 | 11\rangle
          y_{\Jb' 1}(\theta)J_mJ_c\delta
         + \langle\Jb 0 \Jb' 0 | 10\rangle y_{\Jb' 0}(\theta)
          J_m^2
   \right\}.
\end{eqnarray}
Finally, using 
\begin{equation}\label{legendre}
y_{lm}(\theta)=\frac{[l]}{\sqrt{4\pi}}(-1)^m
               \left[\frac{(l-m)!}{(l+m)!}\right]^{1/2}
               P_l^m(\cos\theta),
\end{equation}
one arrives at eqs.~(\ref{94}-\ref{phi-TL'}) 
for the polarized reduced responses in PWIA.



\newpage
{\LARGE {\bf Figure captions}}

\begin{enumerate}

\item Reduced responses at $q=500$ MeV/c for a polarized proton in the 
      $1s_{1/2}$ shell of a mean field potential.
      Solid lines: computed with the CSM; Dashed lines: PWIA (on-shell).

\item As in fig.~1, but now for a proton in the $1p_{1/2}$ or $2s_{1/2}$ shell.

\item Even-rank polarized reduced responses at $q=500$ MeV/c for a proton in 
      the $1p_{3/2}$ shell of a mean field. The meaning of the curves is the 
      same as in fig.~1.

\item The same as fig.~3, but for the odd-rank responses.

\item Even-rank reduced responses for $^{39}$K at $q=500$ MeV/c.
      The meaning of the curves is the same as in fig.~1.

\item The same as fig.~5, but for the odd-rank responses.

\item As in fig.~5, but for $q=300$ MeV/c.

\item As in fig.~6, but for $q=300$ MeV/c.

\item As in fig.~5, but for $q=700$ MeV/c.

\item As in fig.~6, but for $q=700$ MeV/c.

\item Even-rank reduced responses for $^{39}$K at $q=500$ MeV/c.
      Solid lines: CSM; Dashed lines: PWIA (on-shell); Dot-dashed lines: 
      PWIA (CC1). 

\item Even-rank reduced responses for $^{39}$K at $q=500$ MeV/c in the CSM. 
      Solid lines: total responses; Dashed lines: responses including only 
      charge and magnetization terms; Dot-dashed lines: responses including 
      charge, magnetization and convection terms.  

\item The same as fig.~12, but for the odd-rank responses.

\item Response ${W}^{TL}_2$ for $^{39}$K for different 
      values of the momentum transfer. The meaning of the 
      curves is the same as in fig.~12.

\end{enumerate}

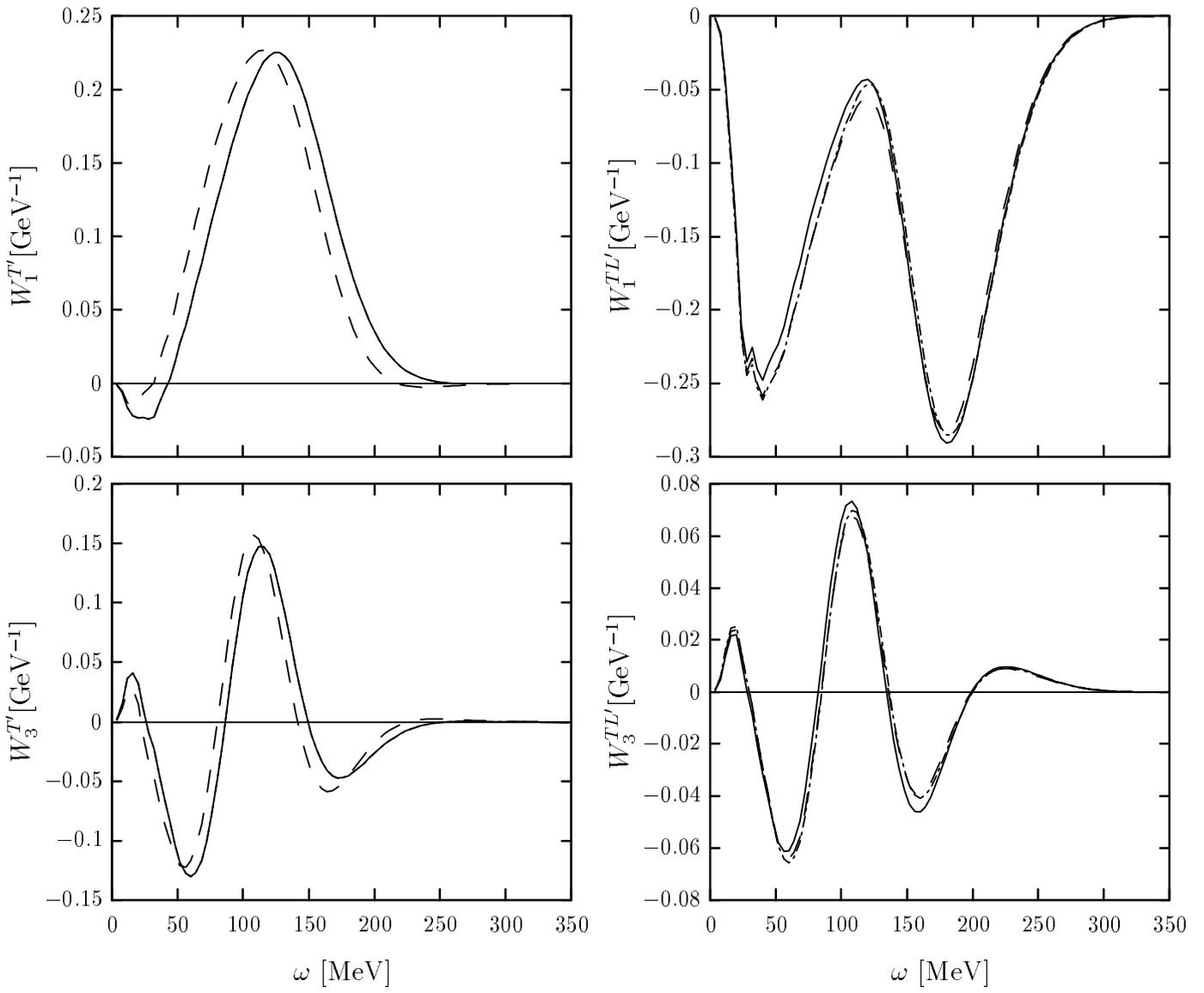
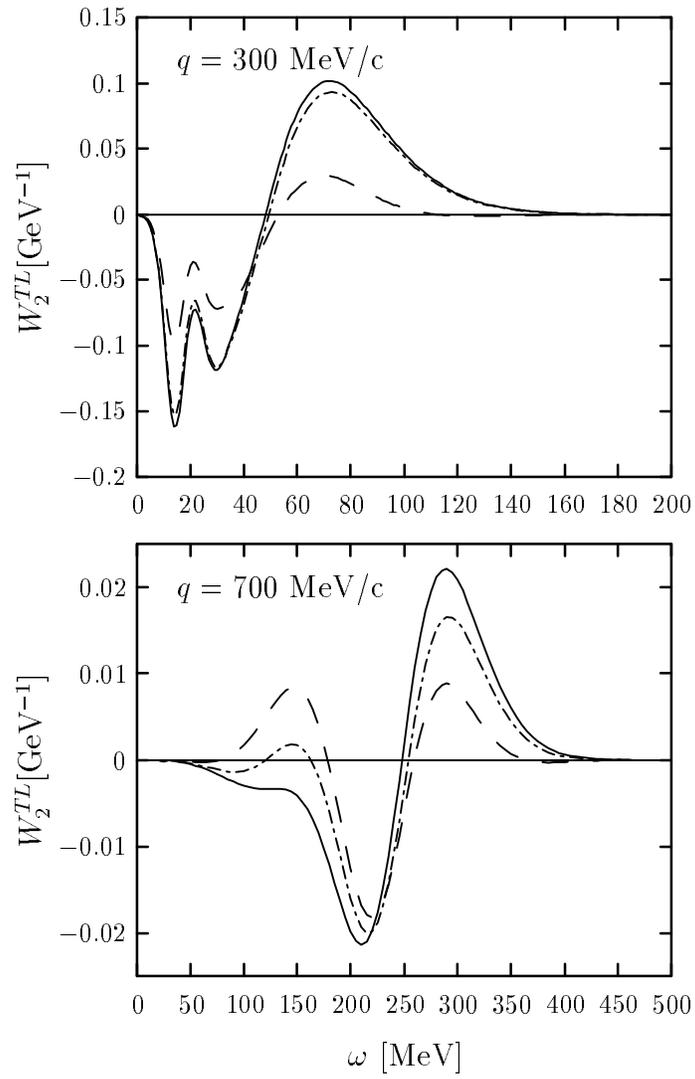
\end{document}